\documentstyle[epsfig]{l-aa}

\def\elogP{$(e, \log P)$}
\def\noy#1#2{\sp{#1}{\rm #2}}

\def\an#1#2#3#4{\noy{#1}{#2}(\alpha,{\rm n})\noy{#3}{#4}}

\def\Msun{M$\sb{\odot}$}

\def\Rsun{R$\sb{\odot}$}
\def\Mbol{M\sb{\rm bol}}
\def\Mv{$M\sb{\rm v}$}

\def\kms{~km~s$\sp{-1}$}

\def\Teff{$T\sb{\rm eff}$}
\def\BD#1#2{BD{$#1$}$^\circ${#2}}
\input epsf

\begin{document}
\title{Insights into the formation of barium and Tc-poor S stars from
an extended sample of orbital elements\thanks{This paper is dedicated
to the memory of Antoine Duquennoy, who contributed many among the
observations used in this study}$^,$\thanks{Based on
observations carried out at the European Southern Observatory (ESO,
La Silla, Chile), and at the 
Swiss telescope at Haute Provence Observatory} }
\author{A. Jorissen\inst{1,2}\thanks{Research Associate, National Fund for 
                                   Scientific Research (FNRS), Belgium}
        \and
        S. Van Eck\inst{1,3}
        \and    
        M. Mayor\inst{3}
\and
        S. Udry\inst{3}
}

\offprints{A. Jorissen (at the address in Belgium)}
\institute{
Institut d'Astronomie et d'Astrophysique,
                  Universit\'e Libre de Bruxelles, C.P.226,
                  Boulevard du Triomphe,
                  B-1050 Bruxelles,
                  Belgium
\and 
Department of Astrophysical Sciences,
Princeton University,
Princeton, New Jersey, U.S.A.
\and
Observatoire de Gen\`eve, CH-1290 Sauverny, Switzerland
}

\date{Received date; accepted date}

\thesaurus{ 08.02.4 -- 08.12.1 }

\maketitle
\markboth{A. Jorissen et al.: Orbital elements of barium and S stars}{A. Jorissen et al.:
Orbital elements of barium and S stars}

\begin{abstract}
The set of orbital elements available for chemically-peculiar red
giant (PRG) stars has been considerably enlarged thanks to a decade-long 
CORAVEL radial-velocity monitoring of 
about 70 barium stars and 50 S stars. 
When account is made for the detection biases, 
the observed binary frequency among strong barium
stars, mild barium stars and
Tc-poor S stars (respectively 35/37, 34/40 and 24/28) 
is compatible with the hypothesis that they are all
members of binary systems. 
The similarity between the orbital-period,
eccentricity and mass-function distributions of Tc-poor S stars and
barium stars confirms that Tc-poor S stars are the cooler analogs
of barium stars.

A comparative analysis of the orbital elements of the various families
of PRG stars, and of a sample of chemically-normal, binary
giants in open clusters, reveals several interesting features.
The eccentricity -- period diagram of PRG stars clearly bears the
signature of dissipative processes associated with mass
transfer, since the maximum
eccentricity observed at a given orbital period is much smaller than in 
the comparison sample of normal giants.
The mass function distribution is compatible with 
the unseen companion being a white dwarf (WD). This lends support to
the scenario of formation of the PRG star by accretion of
heavy-element-rich matter transferred from the
former asymptotic giant branch progenitor of the current WD. 
Assuming that the WD companion
has a mass in the range $0.60\pm0.04$ \Msun, the masses of mild and
strong barium stars amount to $1.9\pm0.2$ and $1.5\pm0.2$ \Msun, respectively.
Mild barium stars are
not restricted to long-period systems, contrarily to what is expected if the
smaller accretion efficiency in wider systems were the dominant
factor controlling the pollution level of the PRG star. 
These results suggest that  
the difference between mild and strong barium stars is mainly one of galactic
population rather than of orbital separation, in agreement with
their respective kinematical properties.

There are indications that metallicity may be the parameter blurring
the period -- Ba-anomaly correlation: at a
given orbital period, increasing levels of heavy-element
overabundances are found in mild barium stars, strong
barium stars, and Pop.II CH stars, corresponding to a sequence of increasingly
older, i.e., more metal-deficient, populations. PRG stars thus seem
to be produced more efficiently in low-metallicity populations. 
Conversely, normal giants in barium-like binary systems may exist in
more metal-rich populations. HD 160538 (DR Dra) may be such an
example, and its very existence indicates at least that binarity is not a
sufficient condition to produce a PRG star.

\end{abstract}

\section{ Introduction }

Barium stars were identified as a class of peculiar red giants by Bidelman
\& Keenan (1951). Typical chemical peculiarities exhibited by these G and K 
giants include overabundances of carbon and of elements heavier than Fe,
like Ba and Sr (e.g. Lambert 1985). These elements bear the signature
of the s-process of nucleosynthesis, a neutron-capture chain starting on Fe
seed nuclei and synthesizing nuclides heavier than Fe located along the valley of
nuclear stability (Burbidge et al. 1957). 
The operation of the s-process is
commonly associated with He-burning thermal pulses occurring on the asymptotic
giant branch (AGB). As a result of the so-called `third dredge-up', 
s-process enriched material is brought to the surface of the AGB star 
(Iben \& Renzini 1983; Sackmann \& Boothroyd 1991). 
Barium stars, as well as their Pop.II counterparts, 
the CH stars (first introduced by Keenan in 1942), are too warm and of too low
a luminosity to have undergone third dredge-ups on the AGB (e.g. Scalo
1976; Bergeat \& Knapik 1997).  
With the discovery
of the binary nature of these stars (McClure et al. 1980; McClure 1983), 
their chemical
peculiarities have been attributed to mass transfer across the binary system. 
When the current WD companion of the barium star was a thermally-pulsing AGB 
star, it transferred s-process- and C-rich material onto its companion, 
which is now viewed as a barium or CH star.

The exact way by which the matter was transferred from the AGB star onto its
companion -- either by wind accretion in a detached binary, or by
Roche lobe overflow (RLOF) in a semi-detached binary --  is still a
matter of debate. On one hand, the fact that barium stars have non-circular
orbits points against RLOF in a semi-detached system, as tidal
effects will efficiently circularize the orbit when the giant is about
to fill its Roche lobe. But on the other hand, 
RLOF seems unavoidable for the barium stars 
with the shortest periods, since their orbital separation is too small  
to have accomodated a large AGB star in a detached binary
system in the past (see in particular the case of HD~121447, having the
second shortest orbital period among barium stars; Jorissen et
al. 1995). The
main difficulty with RLOF is that, when this process involves a giant
star with a convective envelope, it is expected to lead to a dramatic
orbital shrinkage. This so-called `case C' mass transfer is 
dynamically unstable when the mass-loser is the more massive star in
the system. 
A common envelope generally forms at that stage, 
causing a strong drag on the embedded stars (e.g.
Meyer \& Meyer-Hofmeister 1979; Iben \& Tutukov 1993). 
A way out of this dilemna has recently been proposed by Han et al.
(1995), by pointing out that, under  some special conditions, 
the dynamical instability associated with case C mass transfer can be avoided.

The present paper contains an extensive analysis of the orbital
elements of barium stars, since the new orbits presented in
companion papers (Udry et al. 1998ab)
considerably enlarge the database, from the 17 orbits from McClure
\& Woodsworth (1990) that were available to Han et al. (1995) to more
than 50 now.
The binary evolution channels relevant for barium stars, as identified by
Han et al. (1995), will be confronted with our new data, with special
emphasis on the \elogP\ diagram.  
The number of available orbits is now large enough to perform a meaningful
comparison of the period and mass-function
distributions of strong and mild barium stars.  

As far as S stars are concerned, it has become clear that 
Tc-rich and Tc-poor S stars form two separate families with similar  
chemical peculiarities albeit of very different  origins (Iben \& Renzini 
1983; Little et al. 1987; Jorissen \& Mayor 1988; Smith \&
Lambert 1988; Brown et al. 1990; Johnson 1992; Jorissen \& Mayor 1992;
Groenewegen 1993; Johnson et al. 1993; Jorissen et al. 1993; Ake 1997).  
Tc-rich (or `intrinsic') S stars are 
genuine thermally-pulsing AGB stars where the s-process operates in 
relation with the thermal pulses, and where the third dredge-up 
brings the freshly synthesized s-elements (including Tc) to the 
surface (e.g. Iben \& Renzini 1983; Sackmann \& Boothroyd 1991). 
By contrast, Tc-poor  (or 
`extrinsic') S stars are believed to be the cool descendants of barium 
stars. The evolutionary link between barium and S stars is discussed in the
light of the 25 orbits now available for S stars, and appears to be
fully confirmed.

Finally, we present some suggestions to solve the dilemna expressed above 
about the mass transfer mode that operated in barium and extrinsic S stars. 

\section{The stellar samples}
\label{Sect:samples}

Radial-velocity monitoring of several samples of chemically-peculiar
red giants (PRG) has been performed by
the team of McClure at the Dominion Astrophysical Observatory (DAO, Canada)
and by the CORAVEL team on the Swiss 1-m telescope at 
Haute-Provence Observatory (France) and on the Danish 1.54-m telescope
at the European Southern Observatory (La Silla, Chile), with the aim
of deriving their binary frequencies. A detailed
description of the CORAVEL data, along with the  
new orbits, is given in two companion papers (Udry et al. 1998ab; see
Baranne et al. 1979 for a description of the CORAVEL spectro-velocimeter).

A brief description of these samples, on which the present study
relies, is given below, with special emphasis on their
statistical significance. 

\subsection{Barium stars with strong anomalies}

The CORAVEL and DAO samples taken together contain {\it all 34 known barium
stars with strong anomalies} (i.e.  Ba4 or Ba5 on the scale  
defined by Warner 1965) 
from the list of L\"u et al. (1983).
The binary frequency derived for this complete sample in
Sect.~\ref{Sect:frequency} thus allows us to address the question of
whether binarity is a necessary condition to form a strong barium star.
Three stars with a Ba3 index monitored by McClure were included as well in this
sample of strong barium stars.

\subsection{Barium stars with mild anomalies}

The CORAVEL and DAO samples taken together 
include 40 stars with a mild
barium anomaly (Ba$<$1, Ba1 and Ba2 on the scale of Warner 1965). The
CORAVEL sample is a random selection of 33 Ba$<$1,
Ba1 and a few Ba2 stars from the list of
L\"u et al. (1983). Although this sample is by no means
complete, it provides a good comparison to the sample of strong barium
stars described above, for investigating the
correlation between the orbital elements and the intensity of the
chemical anomaly.

Because orbital elements for barium stars are spread in the
literature, Tables~1a and 2a
collect all orbital elements available for
mild and strong barium stars, respectively. The
number in column `Ref.' of these tables refers to the following
papers where the complete set of orbital elements for the considered
star can be found: 0. This paper (see below); 1. Udry et al. (1998a);
2: Udry et al. (1998b);  5. Griffin
(1996);  6. Griffin et al. (1996); 7. Jorissen et al. (1995);
11. Griffin \& Keenan (1992);
12. Griffin (1991); 13. McClure \& Woodsworth (1990);  
20: Griffin \& Griffin (1980). 
Fekel et al. (1993) report preliminary orbital elements for
the mild barium star HD~165141; the lower limit on the orbital period
quoted in Table~1a is derived from their more accurate KPNO data.
For the sake of completeness, a note identifies stars with an orbital
or acceleration solution in the
Hipparcos Double and Multiple Systems Annex (ESA 1997). The comparison
between the astrometric and spectroscopic elements is deferred to a
future study.

For several barium stars monitored by McClure, a few CORAVEL
measurements have been obtained to improve the DAO orbit, since these
measurements significantly increase the time span of the
monitoring. These updated orbits are listed in Table~1a and 2a under
the reference number 0.
A zero-point correction of $-0.46$~\kms\ has been applied to the DAO
measurements, as derived from the average difference in
systemic velocity for the 3 stars (HD 46407, HD 131670 and HD 223617) 
for which independent DAO and CORAVEL orbits are available.  

Several barium stars have very long periods, exceeding the time span of
the monitoring. In those cases, whenever possible, a preliminary orbit
was nevertheless derived by fixing one of the orbital parameters
(usually the period). Those cases can be readily identified in
Tables~1a and 2a by the fact that there is no uncertainty given for the
fixed parameter (see Udry et al. 1998a for more details). 

\subsection {Non-variable S Stars}
\label{Sect:sampleS}

Besides the orbit obtained for the S star HR 1105 (=HD 22649) by
Griffin (1984), our CORAVEL monitoring of a sample 
of 56 S stars is the primary source for investigating the
binary frequency among S stars.
This sample includes 36 bright, northern S stars
from the {\it General Catalogue of
Galactic S Stars} (GCGSS; Stephenson 1984) {\it with no variable star designation},
neither in the {\it General Catalogue of Variable Stars} (Kholopov et
al. 1985) nor in
the {\it New Catalogue of Suspected Variable Stars} (Kukarkin et
al. 1982).  
The criterion of photometric stability has
been adopted to avoid the confusion introduced by the envelope
pulsations masking the radial-velocity variations due to orbital motion. 
Such a selection criterion clearly introduces a strong bias against  
intrinsically bright S stars, which is of importance when deriving the
binary frequency among S stars (see the discussion in
Sect.~\ref{Sect:binaryS}). 

Our samples include the border case HD 121447, sometimes
classified as a Ba5 star and sometimes as an S star; in the analysis
of the orbits presented in the next sections, this star has been included
among {\it both} barium and S stars.  

Table~3a presents all 25 orbits available for
S stars, collected from the following papers, referred to in column
`Ref.' of Table~3a: 0. This paper (see below); 1. Udry et al. (1998a);
3. Carquillat et al. (1998); 7: Jorissen et al. (1995); 
10: Jorissen \& Mayor (1992); 18: Griffin
(1984). The orbits of Jorissen \& Mayor (1992) have been updated with a few new
measurements and listed in Table~3a with reference number 0
in column `Ref'.

\subsection{Mira S stars}

A sample of 13 Mira S stars 
has also been monitored with CORAVEL, 
in order not to restrict the search for binaries to low-luminosity S
stars (see Sect.~\ref{Sect:sampleS}). However, the envelope pulsations
of Mira stars will undoubtedly hamper that search (see
Sect.~\ref{Sect:jitter} and Udry et
al. 1998a for a detailed discussion).  

\subsection{SC and Tc-poor carbon stars}

A sample of 7 SC and CS stars has been monitored as well with CORAVEL,
along with the 3 carbon stars lacking Tc from the list of Little et
al. (1987).

\subsection{CH stars}

Orbits of CH stars are  provided by McClure \& Woodsworth (1990), and
are not repeated here.

\section{The radial-velocity jitter: a new diagnostic}
\label{Sect:jitter}

\begin{figure}
  \vspace{9cm}
  \vskip -0cm
  \begin{picture}(8,8.5)
    \epsfysize=9.8cm
    \epsfxsize=8.5cm
    \epsfbox{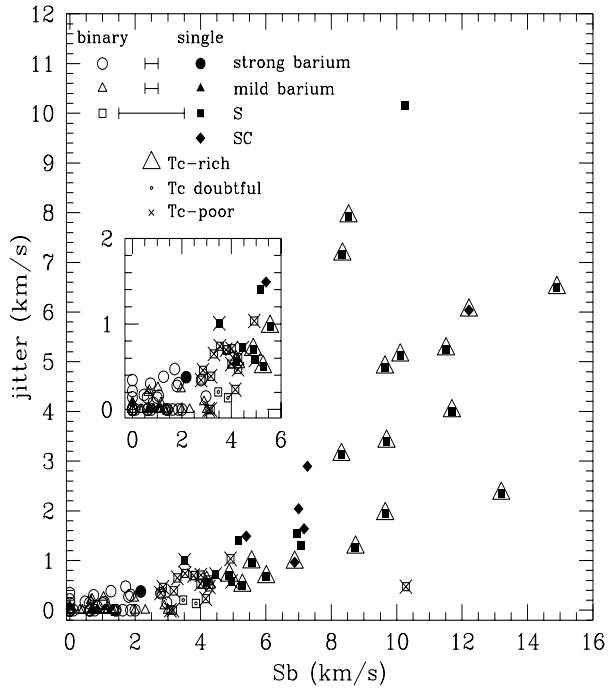}
  \end{picture}
  \vskip -0.5cm
\caption[]{\label{Fig:Sbjitter}
The jitter $(\sigma^2 - \overline{\epsilon}_1^2)^{1/2}$ (where 
$\overline{\epsilon}_1$
is the average uncertainty on one measurement, and $\sigma$ is the standard
deviation of the radial-velocity measurements for non-binary stars, and of the
$O-C$ residuals around the computed orbit for binary stars) 
as a function of the CORAVEL 
line broadening index $Sb$ (see text). Stars with a cc-dip narrower
than the instrumental profile have been assigned $Sb = 0$. Similarly, the jitter
has been set to 0 if $\sigma \le \overline{\epsilon}_1$. Data are from
Tables~1--3. Typical error bars on $Sb$ are displayed in the
upper left corner. 
The suspectedly misclassified S stars HD 262427 and \BD{+22}{4385}
(Table~3b) have not been plotted. The inset is a zoom of the lower
left corner 
}
\end{figure}

The standard deviation of the $O-C$ residuals for some of the orbits computed by
Udry et al. (1998ab) is clearly larger than expected from the error
$\overline{\epsilon}_1$ on one measurement (Tables~1a, 2a and 3a). 
Figure~\ref{Fig:Sbjitter} shows that there is a tendency for the 
largest $O-C$ residuals to be found in the systems with the broadest spectral
lines, as measured by the CORAVEL line broadening index $Sb$. The significance of
this correlation is discussed in this section.
 
The CORAVEL spectrovelocimeter (Baranne et al. 1979) measures the stellar
radial velocity by cross-correlating the stellar spectrum with a mask
reproducing about 1500 lines of neutral and ionized iron-group
species from the spectrum of Arcturus (K1III).
Consequently, the width of the
cross-correlation dip (cc-dip) is an indicator of line broadening.
More precisely, the cc-dip of minor planets (reflecting the sun light),
corrected for the solar rotational velocity and photospheric turbulence,
allows the determination of an `instrumental profile' $\sigma\sb0$.
That parameter corresponds to the sigma of a gaussian function fitted to
the cc-dip of a hypothetical star without rotation and turbulence. An
estimator of the total broadening of stellar spectral lines can then be
derived from the observed width $\sigma$ of the stellar cc-dip as $Sb =
(\sigma\sp2 - \sigma\sb0\sp2)\sp{1/2}$.  

In cool red giants where macroturbulence is the main line-broadening factor, 
the $Sb$ parameter is expected to increase with luminosity, as does
macroturbulence (e.g. Gray 1988). 
This prediction is confirmed from the luminosities derived by Van Eck et al. (1998) from
HIPPARCOS
parallaxes for 23 S stars in common with the present sample. A least-square fit to these data
yields the relation 
\begin{equation}
\label{Eq:MbolSb} 
\Mbol = -1.60 - 0.37 \; Sb,
\end{equation}
valid for $3 \le Sb \le 9$~\kms.
Since bright giants also exhibit large
velocity jitters probably associated with envelope pulsations 
(e.g. Mayor et al. 1984),
a correlation between $Sb$ and the radial velocity
jitter must indeed be expected, as observed on Fig.~\ref{Fig:Sbjitter}. 
Orbits with a
large jitter tend to be associated with giants having 
large $Sb$ indices. This trend is especially clear among binary S
stars, and continues in fact among non-binary S stars (the jitter being then
simply the standard deviation of the radial-velocity measurements). Binary
(`extrinsic') and non-binary (`intrinsic') S stars actually form a continuous
sequence in the ($Sb$, jitter) diagram of Fig.~\ref{Fig:Sbjitter},
the transition between extrinsic and intrinsic S stars
occurring around $Sb = 5$~\kms. Intrinsic S
stars, with their larger $Sb$ indices ($\ga 5$~\kms), may thus be
expected to be more luminous than
extrinsic S stars [$3 \le Sb$ (\kms) $\le 5$]\footnote{The binary S star HDE 332077, with
$Sb=10.3$~\kms,
is outlying in that respect, as in many others (see Sect.~\ref{Sect:fMS})}. This  
conclusion is confirmed by the luminosities derived from the HIPPARCOS parallaxes (Van Eck et
al. 1998).
Extrinsic S stars are in turn more luminous than barium stars, with the border
case HD 121447 (K7IIIBa5 or S0; Keenan 1950, Ake 1979) 
having $Sb = 2.8$~\kms, intermediate between Ba and S stars. 
HD 60197 (K3Ba5) and BD$-14^\circ2678$ (K0Ba1.5) are two other 
barium stars with especially large $Sb$ indices, suggestive of a luminosity larger
than average for barium stars, though there is no information available in the
literature to confirm that suggestion.

Not represented (because accurate $\epsilon_1$ values are lacking) are the two
remarkable mild barium stars HD 77247 ($Sb = 5.8$~\kms) and HD 204075
($Sb = 4.9$~\kms) 
observed at DAO. The latter is indeed known to be a bright giant, with
\Mv\ $ = -1.67$ (Bergeat \& Knapik 1997), thus confirming the  fact  that $Sb$ is a good
luminosity
indicator for red giants.


The large velocity jitter observed in Mira S stars is a consequence of
their complex and variable cc-dips (Barbier et al. 1988; see also
Udry et al. 1998a). 
In some cases however (like AA Cyg and R Hya), the cc-dips are
featureless, broad
and very stable. These stars with a comparatively smaller jitter 
are located on the lower boundary of the region occupied by intrinsic S stars in
Fig.~\ref{Fig:Sbjitter}. 

\section{Binary frequency}
\label{Sect:frequency}

\subsection{Strong barium stars}
\label{Sect:frequencystrong}

According to Table~2, the frequency of binaries
among barium stars with strong Ba indices (Ba3, Ba4 or Ba5) 
is 35/37, the only stars with constant radial-velocities being HD~19014 and HD~65854. 
Before coming to a conclusion as to whether binarity is or is not a necessary 
condition to produce a PRG star, one needs first to assess the barium
nature of the constant stars and second, to evaluate the efficiency
with which binary stars can be detected with our particular protocol
of observations. The latter question is discussed in
Sect.~\ref{Sect:incompleteness}. 

The case of HD~19014 deserves some comments, as its radial velocity appears
definitely variable (Table~2b) though with no clear evidence for binary motion.
Its only distinctive property is its rather large
$Sb$ index of  2.5~\kms (Fig.~\ref{Fig:Sbjitter}), 
suggesting a luminosity larger than
average for barium stars. With a radial-velocity standard deviation of 0.5~\kms,
HD 19014 falls
right on the ($Sb$, jitter) correlation observed in
Fig.~\ref{Fig:Sbjitter}, leaving  no room for variations due to binary motion.

In the absence of any abundance analysis available for HD~19014, the
photometric index $\Delta(38-41)$ defined in the Appendix
may be used instead to infer the level of chemical peculiarities of
that star. Normal
giant stars have $\Delta(38-41) \ge 0$, mild barium stars $-0.1 \le
\Delta(38-41) < 0$, and strong barium stars have $\Delta(38-41) < -0.1$. 
In that respect, HD~19014 appears to have rather weak peculiarities, if any,
as $\Delta(38-41) = -0.01$. Furthermore, in Fig.~\ref{Fig:delta3841},
it lies in between the loci of normal giants and Ib supergiants,
suggesting that the strong barium lines  
are more likely due to a high luminosity  
(as inferred from the large $Sb$ index) than to an
abundance effect. 

HD 65854 is the  other strong barium star with no evidence for binary
motion (McClure \& Woodsworth 1990). Za\v{c}s (1994) performed a
detailed abundance analysis of that star, confirming its barium nature
(see also Table~\ref{Tab:sFeP}). 

Finally, it has to be noted that \BD{+38}{118} is a triple hierarchical system,
with a period ratio $P(ab+c)/P(a+b) =13$ (Table~1a). At this stage, it is not
entirely clear whether the inner pair $a+b$ or the outer pair $ab + c$ is the one
responsible for the barium syndrome.  
Based on the position of the two pairs in the \elogP\ diagram, it is
argued in Sect.~\ref{Sect:elogP} that the barium syndrome is more likely to be associated with
the
wider pair than with the closer one.

\subsection{Mild barium stars}

The risk of misclassifying a supergiant as a mild barium star is high, since 
luminosity also strengthens the very lines of BaII and SrII that are often used
to identify barium stars (Keenan \& Wilson 1977; Smith \& Lambert 1987).  
Excluding the two supergiants obviously misclassified as mild barium stars
and listed in Table~1d,  
the frequency of binary stars among 
barium stars with mild Ba  indices (Ba$<$1, Ba1 and Ba2) 
is at least 34/40 ($= 85$\%; Table~1a), and possibly
37/40 ($= 93$\%) when including the suspected small-amplitude binaries
HD~18182 and HD 183915 (Table~1b), as well as
56 Peg, an interacting binary system (Schindler et al. 1982).
HD~130255 has not been included in the previous
statistics, since it has been shown to be a 
subgiant CH star rather than a barium star
(Lambert et al. 1993). 

Detailed abundance analyses are available for two among the three constant stars
(HD~50843, HD~95345 = 58~Leo and HD~119185).
Sneden et al. (1981) 
find an average overabundance (with respect to solar) of 0.23~dex for
the s-process elements in HD~95345. A similar result is obtained by
McWilliam (1990). The $\Delta(38-41)$ index of 0.07 is compatible with
such a small overabundance level (see the Appendix and Fig.~\ref{Fig:YTi}).
Za\v{c}s et al. (1997) find overabundance levels up to 0.4 dex for s-process
elements in HD~119185.   
No detailed analysis is available for HD~50843, but its 
$\Delta(38-41)$ index of $-0.03$ is comparable to that of HD~119185
($\Delta(38-41) = -0.05$), so that s-process overabundances similar to those of
HD~119185 may be expected for HD~50843.
In summary, all three stars appear to be truly mild barium stars despite the absence of
radial-velocity
variations.

\subsection{S stars}
\label{Sect:binaryS}

The results of the radial velocity monitoring of S stars are presented
in Table~3, which has been subdivided in the following
way: (a) S stars for which an orbit or a lower limit on the orbital
period is available; (b) S
stars with no radial-velocity variations; (c) S stars with radial-velocity
variations but no clear evidence for orbital motion; (d) Mira S stars;
(e) SC stars. 

This partition is motivated by the fact that S stars generally exhibit
some radial-velocity jitter very likely due to envelope pulsation,
that complicates the search for binaries. This is especially true for
Mira S stars and SC stars (see Sect.~\ref{Sect:jitter},
Fig.~\ref{Fig:Sbjitter} and the detailed discussion in Udry et al. 1998a). 
In these conditions, binary  stars are extremely difficult
to find among the stars listed in Tables~3d and e. 
Among these, S UMa may perhaps be binary, but more
measurements are needed before a definite statement can be made. 
The two CS stars with the broadest cc-dips ($Sb > 10$ \kms), R~CMi and RZ~Peg, exhibit
radial-velocity
variations mimicking an orbital motion. These variations are however most probably due to
envelope
pulsations, since their period is identical to the period of the light variations (see Udry
et al. 1998a).
An orbital solution has been found for 
\BD{-08}{1900}, but the binary nature of that star is questionable for several reasons, as
discussed by
Udry et al. (1998a).

The stars listed in Table~3c exhibit 
radial-velocity variations that cannot satisfactorily be fitted by an
orbital solution despite the fact that their jitter is moderate when
compared to that of Mira S stars or SC stars.  

An interesting difference may be noticed between the S stars
with moderate jitter listed in Table~3c and the binary S stars of
Table~3a: as already shown in Sect.~\ref{Sect:jitter}
and Fig.~\ref{Fig:Sbjitter}, binary S stars are restricted to the range
$2.8 \la Sb$ (\kms) $\la 5$, whereas Mira, SC and non-binary S stars
generally have $Sb \ga 4.5$~\kms\ (\BD{-21}{2601}, with $Sb = 3.4$~\kms, is the only exception 
but since
it lacks Tc, it is likely {\it not}
an intrinsic S star).
In Sect.~\ref{Sect:jitter}, it was
argued that this separation reflects a difference in the average luminosities of
these two groups of S stars (Eq.\ref{Eq:MbolSb}), as confirmed since by Van Eck et al. (1998). 
This is well in
line with the conclusion of previous studies (e.g. Brown et al. 1990; Johnson 1992;
Jorissen et al. 1993) that two distinct families,
having a very different evolution history, are found among S stars. 
This `binary paradigm' 
states that all Tc-poor S stars should be binaries, being the cool descendants 
of the barium stars, whereas Tc-rich S stars are genuine
thermally-pulsing AGB stars and
ought not be binaries. 
The data presented in Table~3 largely confirm that
paradigm, since all Tc-poor S stars (with the only exceptions of HD
189581 and \BD{-21}{2601}) are binary stars. The presence of Tc-rich S stars among the
binary stars, although allowed in principle, is limited to HD 63733,
HD 170970 and $o^1$ Ori. As argued by Jorissen et al. (1993), the
presence of Tc in the former two stars is even questionable, since they
fall on the boundary between Tc-rich and Tc-poor stars according to
the criterion of Smith \& Lambert (1988). 
The absence of infrared excesses (Jorissen et al.
1993) and their small $Sb$ indices (Fig.~\ref{Fig:Sbjitter}, where HD 63733 and
HD 170970 are flagged as `Tc doubtful') lend support to their extrinsic
nature. 
$o^1$ Ori is a very special  
case, as it shares the properties of extrinsic (in having a WD
companion; Ake \& Johnson 1988) and intrinsic (in having Tc) S
stars. It may be an extrinsic S stars starting its ascent on the thermally-pulsing AGB.  

Finally, it should be noted that two S stars with constant
radial velocities (Table~3b) were found. The absence of
any detectable jitter and their small $Sb$ index are, however, quite unusual for
S stars. We therefore suspect that these stars may have been
misclassified as S stars. This suspicion appears justified at least
for HD~262427, which is listed as S? in the discovery
paper of Perraud (1959).     

\subsection{Tc-poor carbon stars}

In their extensive study of Tc in late-type stars, Little et
al. (1987) list only 3 carbon stars lacking Tc lines (X Cnc, SS Vir
and UU Aur). 
They might possibly be the analogs 
of the extrinsic, Tc-poor S stars. If so, they should be
binary stars. The results of their CORAVEL
monitoring is presented in Table~4, with no clear evidence for binary
motion, except perhaps in the case of X Cnc. The large jitter
exhibited by these stars (especially SS Vir) is reminiscent of the
situation encountered for SC stars, and like them, the Tc-poor C stars have large
$B-V$ indices ($\ge 3.0$).   

\section{Incompleteness study}
\label{Sect:incompleteness} 

\subsection{General principles}
\label{Sect:MonteCarlo}

In order to evaluate the real frequency of binary stars within a given stellar
sample, it is of key importance to properly evaluate the efficiency with 
which binary systems can be detected. That detection
efficiency not only depends
upon {\it internal} factors set by the protocol of observations (internal velocity
error, sampling and time span
of the observations), but also upon {\it external} factors related to 
the orbital properties of the binary systems [through the distributions of periods
$P$, eccentricities $e$, and ratios 
$Q = M_2^3/(M_1+M_2)^2$], 
or to their orientation with respect to the line of sight (through the
inclinations $i$ and  
longitudes of periastron $\omega$), or with respect to the time sampling (through
the epochs $T$ of passage at periastron).

A Monte-Carlo simulation of our ability to detect binary stars 
has therefore been performed following the guidelines described by
Duquennoy \& Mayor (1991).
First, $N$ binary systems (where $N$ is the number of stars in the
observed sample being tested) are
generated by drawing $T_j$ and $\omega_j$ ($1\le j \le N$)  
from uniform random distributions, $i_j$ from a $\sin i$
probability distribution (implying random orientation of the orbital poles on the
sky), and $P_j$, $e_j$ and 
$Q_j$ from their observed distributions extrapolated in several different ways (see
Sect.~\ref{Sect:PeQ}). 

The synthetic binary $j$ 
is then attributed the line-width index
$Sb_j$ of the $j$th star in the real sample, and similarly a set of
observation dates $t_{j,k}$ ($1 \le k \le n^{\rm obs}_j$, $n^{\rm obs}_j$ being
the number of observations of the real star $j$) and a set of internal velocity
errors $\epsilon^{\rm int}_{j,k}$. 
The intrinsic radial-velocity jitter observed for S stars
(and, to a much lesser extent, for barium stars; see Fig.~\ref{Fig:Sbjitter} and
Sect.~\ref{Sect:jitter}) has to be included in the simulation.
To that purpose, a parabolic fit to the trend observed in Fig.~\ref{Fig:Sbjitter}
has been used to associate a radial-velocity jitter $\epsilon^{\rm jit}_j$ to the
selected line width $Sb_j$.
Synthetic velocities $V_{j,k}$ are then computed for the observation dates
$t_{j,k}$ from the orbital elements, with an added error drawn 
from a gaussian distribution of standard deviation 
$\epsilon_{j}=[(\overline{\epsilon^{\rm int}_j})^2+(\epsilon^{\rm
jit}_j)^2]^{1/2}$, $\overline{\epsilon^{\rm int}_j}$ being the average internal
error on one measurement of star $j$.
The few stars 
measured by other authors\footnote{HD~165141 has not been included 
in the incompleteness study described in this section,
because it was added to Table~1a later on}
have been 
attributed sets of observation dates, internal velocity errors, line-width indices
and jitter from stars of our sample having comparable orbital elements.

Finally, the binary star $j$ is flagged as detected if $P_{\nu}(\chi^2_j)<0.01$,
where $P_{\nu}(\chi^2)$ is the $\chi^2$ probability function
with $\nu=n^{\rm obs}_j - 1$ degrees of freedom, and $\chi^2_j=(n^{\rm obs}_j -
1)(\sigma_j/\epsilon_j)^2$, $\sigma_j$ being the unbiased
dispersion of the $n^{\rm obs}_j$ synthetic velocities of star $j$. The
detection rate is then $N_{\rm bin}/N$, where $N_{\rm bin}$ is the number of stars
with $P_{\nu}(\chi^2_j)<0.01$.

This procedure is repeated until 100 sets of $N$ stars have been generated, which
is sufficient for the {\it average} detection rate to reach an asymptotic value.
The above method has been applied separately for the sample of 37 barium stars
with strong anomalies, of 40 mild barium stars and of 28 S stars with $Sb < 5$~\kms\ 
(referred to as non-Mira S stars in the following).
This particular choice for the $Sb$ threshold is motivated by the fact
that on Fig.~\ref{Fig:Sbjitter}, Tc-poor S stars (that are expected to belong to
binary systems) are restricted to $Sb < 5$~\kms. 
The method has not been applied to Mira S stars because of the
uncertain (and probably large) radial-velocity jitter affecting these stars 
(Sects.~\ref{Sect:jitter}, \ref{Sect:MCdiscussion} and Fig.~\ref{Fig:Sbjitter}).

\subsection{Detection rates for specific $P$, $e$ and $Q$ distributions}
\label{Sect:PeQ}

The main difficulty of the Monte-Carlo method outlined in
Sect.~\ref{Sect:MonteCarlo} is that 
the {\it real} distributions of period, eccentricity and
$Q$  are not completely known,
since the very detection biases we want to evaluate render the {\it observed}
distributions incomplete. Different choices have therefore been made on how to
complete the observed distributions, and the sensitivity of the estimated binary
detection rates on these choices is evaluated {\it a posteriori}. 

\begin{figure*}
  \vskip -0cm
\vspace{13cm}
  \begin{picture}(18,9)
    \epsfysize=13cm
    \epsfxsize=18cm
    \epsfbox{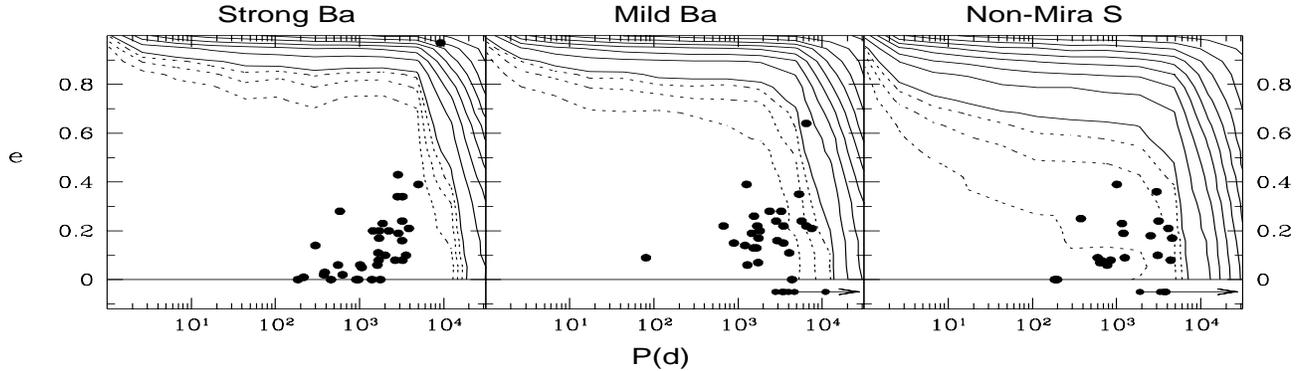}
  \end{picture}
  \vskip -7cm
\caption[]{\label{Fig:isoproba}
The iso-probability curves in the \elogP\ plane 
for detecting binary systems among strong barium, mild barium and non-Mira S stars.
The dotted curves correspond
to detection probabilities of 97.5\%, 95\% and 92.5\% (decreasing
towards the upper right corner),
and the solid curves to probabilities of 90\%, 80\%,
70\%,... 
In the lower part of each diagram are represented binary stars for which
only a lower limit is available on the period
}
\end{figure*}

The distribution of ratios $Q$ 
is likely to have little impact on the detection biases. 
Therefore, the observed distribution as derived in
Sect.~\ref{Sect:fM} has been adopted. 
Three different cases are considered
for the period and eccentricity distributions, as follows:
\medskip\\
\noindent {\it (i) Uniform distribution in the \elogP\ diagram}
\label{Sect:uniform}
\smallskip\\
Figure~\ref{Fig:isoproba} shows the curves of iso-probability detection
for strong barium, mild barium and non-Mira S stars in the \elogP\ plane.
To derive these probabilities,
the \elogP\ plane has been uniformly covered by a mesh of 480 points,
which is equivalent to adopting uniform $\log P$ and $e$ distributions.
For each of these mesh points, $100 N$ (i.e. about 3000) synthetic
binaries have been
generated as indicated above, in order that the detection probability
be mainly set by $P$ and $e$ rather than $i, T$ and $\omega$.     
As seen on Fig.~\ref{Fig:isoproba}, the detection probability drops
tremendously for  $P \ga  8000$~d,
because of the finite timespan of the observations,
which started in 1985 for most strong barium stars, and in 1986
for most mild barium and S stars. For S stars, the intrinsic jitter affecting
their radial velocities also contributes to lower the detection rate.
\medskip\\

\noindent{\it (ii) Observed $P$ and $e$ distributions}
\smallskip\\
In this case, the observed $P$ and $e$ distributions are assumed to 
represent the {\it real} distributions, as if there were no systems with periods
longer or eccentricities larger than those currently detected in the real samples.
This is a conservative choice that allows one 
to estimate the {\it maximum} detection rate. 
With this crude hypothesis, an unfavorable
spatial or temporal orientation of the binary system is the only possible cause
of non-detection. 
Eccentricities and periods of the synthetic binaries were drawn in accordance with
their observed distribution in the \elogP\ diagram, using the rejection method
described by Press et al. (1992). 

The binary detection rates obtained for the three samples under these
hypotheses are listed in Table~\ref{Tab:MonteCarlo} under item
Monte-Carlo (ii).
\medskip\\  
\noindent {\it (iii) Extrapolated $P$ and $e$ distributions}
\smallskip\\
In this last case, the real period distribution is assumed to be identical to the
observed distribution at short periods. Its long-period tail includes the $N_{\rm
no-P}$ stars having no orbital period currently available (i.e. stars with only
a lower limit available on $P$, or even stars with constant
radial velocities) but assumed to be very
long-period binaries. More precisely, the  $N_{\rm no-P}$ stars are 
uniformly redistributed over the period bins ranging from $P_{\rm
inf}$ to 20000 d,
where $P_{\rm inf}$ is the first period bin containing a star from
the $N_{\rm no-P}$ subsample.
Eccentricities are supposed to be uniformly distributed under the curve $e= (\log P-1)^2/7$.

The corresponding binary detection rates are listed in
Table~\ref{Tab:MonteCarlo} 
under item Monte-Carlo (iii).
\medskip\\
What are the main causes of non-detection? 
In the case of mild barium stars for example, the non-detected
binaries in simulation (iii) are distributed as follows,  in
decreasing order of importance:  
$e > 0.8$      for 47.3\%,
$P > 10^4$d for 32.7\%,
$-\pi/10 < \omega ({\rm mod} \pi) <\pi/10$ for 20.7\% 
(because the radial-velocity curve remains very flat 
over a large fraction of the orbital cycle around apastron),
$\sin i < 0.1$ for 4.5\%, 
and a combination of various less severe conditions for 9.8\%. 
Note that, since a star may belong to more than one of these categories, the sum
of the above percentages exceeds 100\%.
In more practical terms, these numbers translate into 6.6\% of the undetected
binaries having a radial-velocity semi-amplitude $K < 0.4$~\kms\ too
small in comparison with
the instrumental accuracy, while 37.8\% remained undetected because of
an incomplete phase coverage ($\Delta t < P/4$) and 26.6\% because the number
of measurements is too small ($<8$).

\subsection{Discussion}
\label{Sect:MCdiscussion}

The detection rates obtained for the various cases considered in
Sect.~\ref{Sect:PeQ} are summarized in Table~\ref{Tab:MonteCarlo}, where they are
compared with the observed rates.

\setcounter{table}{5}
\begin{table*}
\caption[]{\label{Tab:MonteCarlo}
Comparison of the simulated and observed binary detection rates for strong barium
stars, mild barium stars and non-Mira S stars ($Sb < 5$~\kms). 
The quoted uncertainty corresponds to the standard deviation
of the detection rates 
for the 100 stochastically independent sets generated
} 
\begin{tabular}{lccc}
          &  strong Ba    & mild Ba  & S ($Sb < 5$ \kms) \cr
\hline
Observed  & 94.6\% = 35/37&    85.0\% = 34/40 & 85.7\%   \\
          &               & to 92.5\% = 37/40 & = 24/28  \\
\smallskip\cr
Monte-Carlo (ii) & 97.9 $\pm$ 3.0\%      & 93.0 $\pm$ 5.0\%      & 96.8 $\pm$ 4.0\%      \cr
                 & =(36.2 $\pm$ 1.1)/37 & =(37.2 $\pm$ 2.0)/40 & =(27.1 $\pm$ 1.1)/28 \cr
\smallskip\cr
Monte-Carlo (iii)& 95.7 $\pm$ 4.0\%      & 89.4 $\pm$ 6.0\%      & 89.5 $\pm$ 6.0\%      \cr
                 & =(35.4 $\pm$ 1.5)/37 & =(35.8 $\pm$ 2.4)/40 & =(25.0 $\pm$ 1.5)/28 \cr
\hline
\end{tabular}
\end{table*}

If binarity were not the rule among barium stars, the observed rates of binaries
would be significantly lower than the predicted ones. 
Here on the contrary, there is a good agreement between the Monte-Carlo 
predictions and the observed binary rates among mild and strong
Ba stars. One may therefore conclude that {\it binarity is a necessary condition
to produce chemically-peculiar red giants like mild and strong Ba stars}.
The agreement is slightly better for the Monte-Carlo simulation 
(iii) extrapolating
the period distribution up to 20000~d. The few barium stars
with undetected radial-velocity variations are thus likely 
binaries with very long periods or 
with unfavourably-oriented orbits. 

The situation is more tricky for S stars.  
The application of the Monte-Carlo method to the {\it whole} sample of S
stars is hampered by our ignorance of the exact amount of jitter
in the radial velocities of Mira stars with large $Sb$ indices (see
Fig.~\ref{Fig:Sbjitter}).   
Therefore the Monte-Carlo simulation cannot be used to evaluate the rate of
binaries among the whole sample of S stars (as we did for barium stars).
The problem is less severe for S stars with $Sb < 5$ \kms, where the
jitter is smaller ($< 1$ \kms) and may be estimated with more
confidence. Moreover, the fact that Tc-poor S stars (i.e. extrinsic S
stars suspected of belonging to binary systems) appear to be
restricted to  $Sb < 5$ \kms\ provides an independent justification
for applying the Monte-Carlo simulation to that subsample.
The observed
binary rate for S stars with $Sb < 5$ \kms\ is close to the one predicted with the
extrapolated $P$ and $e$ distributions (case iii in
Table~\ref{Tab:MonteCarlo}).  
This result is thus
consistent with the hypothesis that all S stars with $Sb < 5$~\kms\
are members of binary systems [Note that, for S stars, the case (iii)
predictions
have to be preferred over the case (ii) ones, since  
the period distribution of S stars is most probably incomplete at large
periods, due to a more limited time coverage than in the case of barium stars]. 
However, this agreement should not be overinterpreted, as the $Sb <
5$~\kms\ limit between extrinsic and intrinsic S stars may be somewhat
fuzzy.

\begin{figure}
\vspace{9cm}
  \vskip -0cm
  \begin{picture}(8,8.5)
    \epsfysize=9.8cm
    \epsfxsize=8.5cm
    \epsfbox{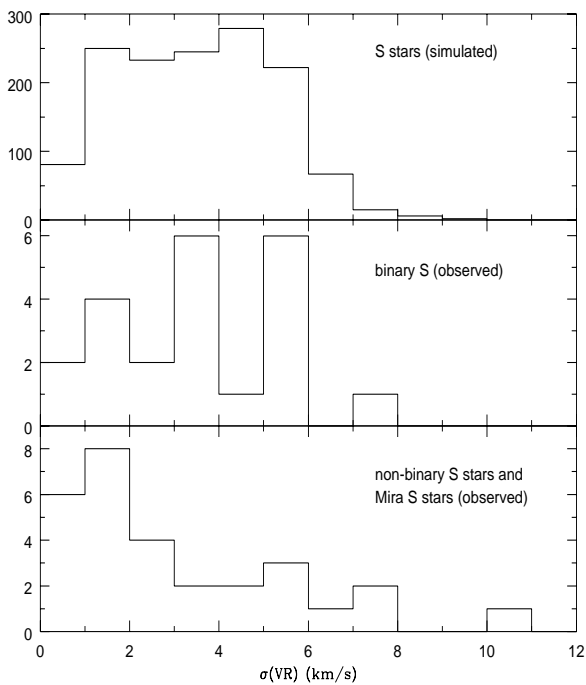}
  \end{picture}
  \vskip -0.5cm
\caption[]{\label{Fig:histosigvr}
The {\it simulated} $\sigma(V_r)$ distribution for binary S stars (upper panel),
and the {\it observed} distribution for real S stars (lower two panels).
HD 191589 and HD 332077, the two binary S stars with very discrepant mass functions,
lie outside the boundaries of the middle panel, 
with $\sigma(V_r)$=14.4 and 18.5 \kms\ respectively
}
\end{figure}

Are there binaries in our sample of intrinsic S stars? For these
stars, the intrinsic
jitter may possibly induce variations of the same order
of magnitude as those caused by binarity, and thus renders the detection 
of possible binaries very delicate.
However, several arguments indicate that binary stars are probably
not very frequent among the intrinsic S stars of our sample.
The upper panel of Fig.~\ref{Fig:histosigvr} shows the distribution of
the radial-velocities standard deviation $\sigma(V_r)$
{\it predicted} by the Monte-Carlo simulation
for a sample of binary stars having orbital parameters matching those
of the binary S stars, and with 
an intrinsic jitter of 1~\kms.
As expected, the {\it observed} $\sigma(V_r)$ distribution for 
binary S stars (middle panel of Fig.~\ref{Fig:histosigvr})  
matches the simulated distribution 
(allowing for large statistical fluctuations due
to the small number of stars observed). 
On the contrary, the observed $\sigma(V_r)$ distribution for
Mira S stars (lower panel of Fig.~\ref{Fig:histosigvr}) differs
markedly from that of binary S stars, since the former distribution peaks  at 
$\sigma(V_r)\sim 1.5$\kms\ and is rapidly falling off at larger $\sigma(V_r)$.
The paucity of Mira S stars with $2 \le \sigma(V_r) \le 6$~\kms\ 
must therefore reflect the low percentage of binaries among this group.
The minimum value of the jitter (1\kms) adopted in the simulation
is a conservative choice;  a larger jitter would shift the simulated
$\sigma(V_r)$ distribution (upper panel) towards larger $\sigma(V_r)$
values,  thus strengthening the above conclusion.

Although binarity is not required to produce intrinsic S stars and is
indeed not frequent in the sample considered in this paper, binary stars may
nevertheless exist among them as in any class of stars. 
A few intrinsic S stars with main sequence companions are known, from the composite nature of
their
spectrum at minimum light. 
They include T Sgr, W Aql, 
WY Cas (Herbig 1965; Culver \& Ianna 1975), and
possibly S Lyr (Merrill 1956), as well as the close visual binary 
$\pi^1$ Gru (Feast 1953). 
A Tc-rich S star with a WD companion is also known ($o^1$~Ori; Ake \& Johnson 1988).



\section{The \elogP\ diagram}
\label{Sect:elogP}

\begin{figure}
\vspace{9cm}
  \vskip -0cm
  \begin{picture}(8,8.5)
    \epsfysize=9.8cm
    \epsfxsize=8.5cm
    \epsfbox{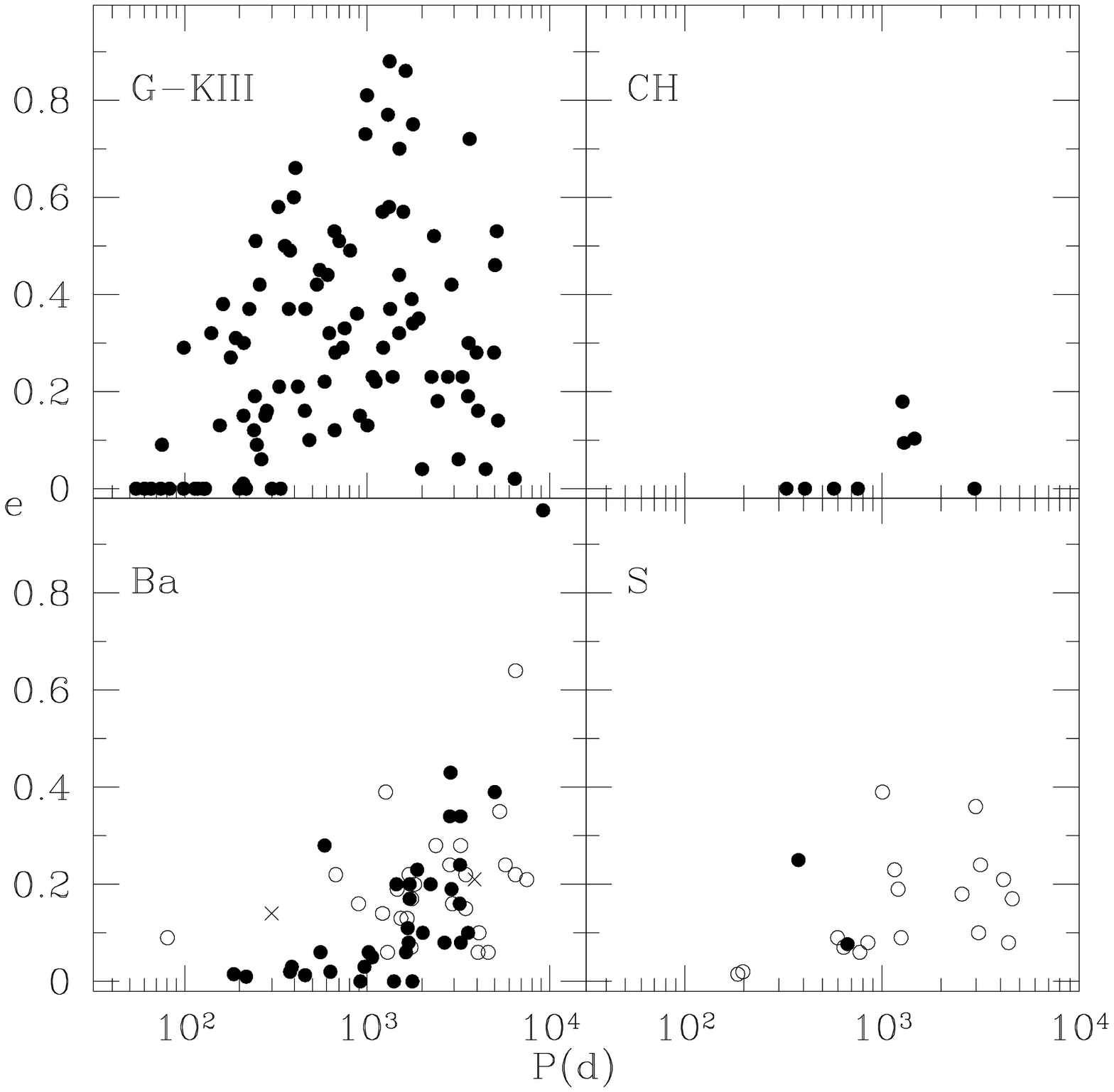}
  \end{picture}
  \vskip -0.5cm
\caption[]{\label{Fig:elogP}
The \elogP\ diagram for various samples of red giant stars:\\
Lower left panel: Barium stars. Mild (Ba$<$1 -- Ba2, from
Table~1a) and strong (Ba3 -- Ba5, from
Table~2a) barium stars are represented by open and filled
circles, respectively. 
The crosses identify the two pairs of the triple hierarchical system
\BD{+38}{118};\\
Upper left panel: Binaries involving G and K giants in open clusters (Mermilliod
1996);\\
Upper right panel: CH stars (McClure \& Woodsworth 1990);\\
Lower right panel: S stars from Table~3a
(Note that HD~121447, the border case
between barium and S stars, has been included in both samples).
The filled circles correspond to
HD~191589 and HD~332077, two S stars with unusually large mass
functions (see Sect.~\protect\ref{Sect:fM})
}
\end{figure}

The \elogP\ diagram is a very useful tool to study binary evolution,
as the various processes modifying the orbital parameters in the
course of the evolution (like tidal interaction, RLOF or wind
accretion) imprint distinctive signatures on the \elogP\ diagram (see the
various papers in {\it Binaries as tracers of stellar
formation}, edited by Duquennoy \& Mayor 1992). 
The \elogP\ diagram for various classes of red giants of interest here
is displayed in Fig.~\ref{Fig:elogP}. 


The sample of G and K giants from open clusters presented in
Fig.~\ref{Fig:elogP} (from Mermilliod 1996) will be used as a reference sample to
which the  binaries involving PRG stars may be compared.
A striking feature of this \elogP\
diagram is the relative paucity of
systems with $e < 0.1$ and orbital periods longer than 350~d. A similar lack of
circular systems is observed among binary systems involving solar-type main-sequence primaries
(Duquennoy \&
Mayor 1991; Duquennoy et al. 1992), the threshold period (10~d)
being much shorter in this case. 
On the contrary, many systems with periods shorter than this threshold have
circular orbits.

Both features are the result of physical processes that operated in the former
history of these systems. Tidal effects on the more evolved component nearly
filling its Roche lobe are responsible for the circularization of the closest
binaries in a given (coeval) sample. The threshold period is then set
by the  largest radius
reached by the more evolved component in its former evolution (Duquennoy et
al. 1992; 
Mermilliod \& Mayor 1996). The lack of circular systems above the tidal
circularization threshold has been interpreted as an indication that binary
systems form in eccentric orbits. A theoretical support to this hypothesis is
provided by Lubow \& Artymowicz (1992). These authors show that 
the interaction betwen the young binary system  and a circumbinary disk
containing proto-stellar residual material may increase
a moderate initial eccentricity, thus giving rise to the observed lack of circular
orbits among unevolved systems.

The \elogP\ diagram of barium stars is markedly different from that of
cluster giants, for (i) 
barium stars nearly fill the low-eccentricity gap observed among unevolved
binaries\footnote{The newly derived orbits for barium systems 
were never forced to be circular, even though the criterion of Lucy \&
Sweeney (1971) may indicate that the data is compatible with the
hypothesis $e=0$ at the 5\% level. Given the general appearance of the
\elogP\ diagram (Fig.~\protect\ref{Fig:elogP}), there is no physical
reason not to accept small albeit non-zero eccentricities}, 
(ii) the minimum and maximum periods for a circular orbit are  $P_1 \sim
200$~d and $P_2 \sim 4400$~d, respectively, among barium stars, as compared to 50
and 350~d for cluster giants, and (iii) at a given
orbital period, the maximum
eccentricity found among barium systems is much smaller than for cluster
giants. Still, it is important to note that quite 
large eccentricities ($e \sim 0.97$, HD~123949) are found among barium
stars, yet at large periods ($P \sim 9200$~d).
The \elogP\ diagrams of S and CH stars, and their threshold periods $P_1$ and
$P_2$ in particular, are very similar to those of barium stars.

The differences between the \elogP\ diagrams of PRG and cluster
binaries reflect 
the fact that the orbits of PRG systems have been shaped by the mass-transfer
process responsible for their chemical peculiarities, whereas most of the cluster
binaries are probably pre-mass transfer binaries.  
The \elogP\ diagram of PRG stars may also have been altered to some
extent by tidal effects occurring in more recent phases (e.g., when
low-mass barium stars currently in the clump evolved up the RGB), thus
complicating its interpretation in terms of mass-transfer only (see
Sect.~\ref{Sect:Pdist}).  
 Detailed models of binary evolution are therefore required to
fully interpret the \elogP\ diagram
of PRG stars. They are deferred to a forthcoming paper.



\begin{figure}
\vspace{9cm}
  \vskip -0cm
  \begin{picture}(8,8.5)
    \epsfysize=9.8cm
    \epsfxsize=8.5cm
    \epsfbox{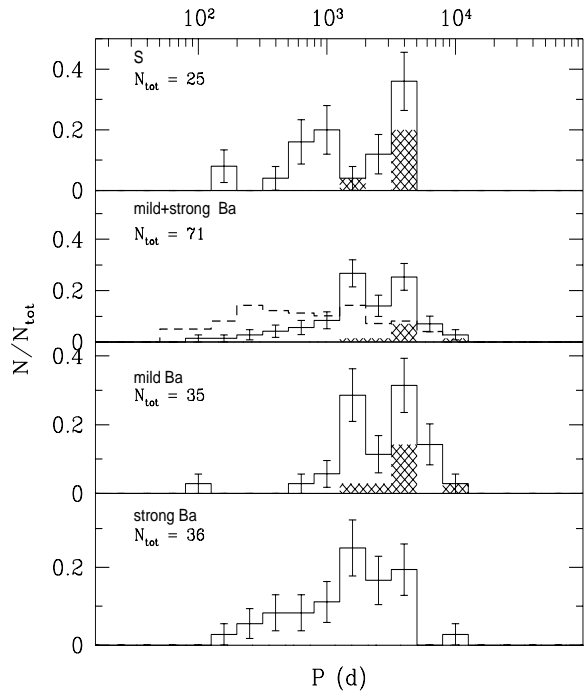}
  \end{picture}
  \vskip -0.5cm
\caption[]{\label{Fig:PS}
Period distributions for various families of PRG stars. Shaded regions
in the histograms denote systems with only a lower limit
available on the orbital period. For comparison, 
the thick dashed line in the mild+strong Ba
panel provides the period distribution for the sample of normal giants
in open clusters (Mermilliod 1996; see
Fig.~\protect\ref{Fig:elogP}). The error bars on the histograms
correspond to the statistical error expected for a Poisson distribution 
}
\end{figure}

Two barium stars deserve some comments. 
As already indicated in Sect.~\ref{Sect:frequencystrong}, \BD{+38}{118} is  a
triple system. The large eccentricity ($e = 0.14$) of the inner pair ($P = 300$~d)
may therefore not be representative of barium systems and should probably not be
regarded as
a constraint on the mass-transfer process that shaped the barium-star orbits. 
Mazeh \& Shaham (1979) and Mazeh (1990) have shown that, in a triple system, the
dynamical interaction of the inner binary with the third body 
prevents the total circularization of the inner orbit. However, the oscillation
of the eccentricity of the inner binary obtained in the cases considered by these
authors has an amplitude (of the order of 0.05) much smaller than the current
eccentricity of \BD{+38}{118}. It is not clear therefore whether such a dynamical
interaction in a triple system may be responsible for the large eccentricity of
the inner pair in \BD{+38}{118}. Another possibility is that the mass transfer
responsible for the barium syndrome actually originated from the distant third
companion, whereas the inner pair consisting of the barium giant and a low-mass
main sequence companion has orbital elements typical of unevolved systems like
those involving giants in clusters. The mass functions of the two pairs are
similar (see Sect.~\ref{Sect:fM}) and do not contradict the above statement,
although they cannot be used to confirm it either.

Another remarkable system is HD~77247, with $P = 80.5$~d and $e =
0.09\pm0.01$. There is no indication
whatsoever from the $O-C$ residuals of the orbit of McClure \&
Woodsworth (1990) that this star may belong to a triple system. 
Yet its orbit should have been circularized by the tidal processes at
work in giants, unless that star is much younger than
the cluster giants displayed in Fig.~\ref{Fig:elogP} (see Duquennoy et
al. 1992).  An interesting property in that respect is the fact that
the star has anomalously broad spectral lines ($Sb = 5.8$~\kms;
Table~1a), an indication that it is either a luminous G (super)giant
or that it is rapidly rotating.

\section{Period distributions}
\label{Sect:Pdist}

\begin{table*}
\caption[]{\label{Tab:Pdist}
Summary of the distinctive features of the period distributions of mild and 
strong barium stars
}
\begin{tabular}{lcc}
&  \multicolumn{1}{c}{Strong Ba}             & \multicolumn{1}{c}{Mild Ba}    \cr
\hline\cr
$\bullet$ short-period mode ($P \la 1500$~d) & $e< 0.08$
          & no short-period mode \cr
          & & (except for the peculiar system HD 77247)
\medskip\cr
$\bullet$ long-period mode ($P \ga 1500$~d) &  \multicolumn{2}{c}{mild and strong barium stars
in this mode  are indistinguishable}\cr
         & \multicolumn{2}{c}{$e> 0.08$}
\medskip\cr
$\bullet$ upper-period cutoff & 5000~d & $> 11000$~d \cr
           & (except for the special case HD 123949:\cr
           &  $P \sim 9200$~d, $e = 0.97$)\cr 
\hline
\end{tabular}
\end{table*}

The orbital-period distributions of mild barium stars, strong barium stars and S
stars are compared in Fig.~\ref{Fig:PS}, and their distinctive
features are outlined in Table~\ref{Tab:Pdist}. The significance of
the two modes identified in the period and eccentricity distributions of strong barium
stars may be grasped by considering the Roche radii 
corresponding to the mode boundaries. Adopting 2.1 \Msun\ as the typical
mass of strong-barium systems (1.5 + 0.6 \Msun; see Table~\ref{Tab:fM}), 
the $\sim 1500$~d
threshold between the low- and long-period modes translates into
$A \sim 700$~\Rsun\ and $R_{\rm Roche} \sim 200$~\Rsun\ for
the Roche radius around the former AGB companion in its final stage
(when its mass amounts to $\sim 0.6$~\Msun). Since this Roche radius is of the
order of AGB radii, the threshold between the short- and
long-period modes is probably related to the different mass transfer
modes arising in detached and semi-detached binary systems (see
Sect.~\ref{Sect:RLOF}). The lower boundary of the
short-period mode ($\sim 200$~d) corresponds to $A \sim
180$~\Rsun, or   $R_{\rm Roche} \sim 85$~\Rsun\ around the barium
star. Since this Roche radius is of the order of radii reached on the
RGB, the lower end of the short-period mode is likely altered by tidal
circularization or even RLOF occurring as the current barium star
evolves on the RGB (see Sect.~\ref{Sect:S}). 

The upper period cutoff for strong
barium stars (which is meaningful, since the sample is complete) 
is significantly smaller than that of mild barium stars, as
expected in the framework of mass transfer through wind accretion
(Boffin \& Jorissen 1988; Jorissen \& Boffin 1992;
Sect.~\ref{Sect:fMpop}).
The level of chemical
peculiarities of a barium star depends, among other parameters, 
on the amount of matter transferred onto it. 
Since that amount may in turn be expected to be
smaller in wider (i.e. longer-period)\footnote{It is in fact the
periastron distance rather than the
orbital period which is the key parameter in this respect. The
long period ($\sim 9200$~d) observed for the strong barium star HD 123949 is
therefore not relevant, since its very large eccentricity yields a
periastron distance much smaller than in systems with $P \sim 5000$~d and
smaller eccentricities}
systems, it is not surprising
that the long-period cutoff is larger for mild barium
stars. The period cutoffs observed for strong and mild barium stars
(5000 and $> 11000$ d respectively) therefore put constraints on the
efficiency of wind accretion, to be used in future simulations. 

More generally,
milder chemical peculiarities are expected in longer-period systems.
However, the broad overlap between the period distributions of mild and strong
barium stars (Fig.~\ref{Fig:PS}) suggests that the scatter in a
(period, chemical anomaly) diagram will be large. 
It is therefore likely that other parameters play an
important role in controlling the level of chemical
peculiarities. That question
will be addressed in Sects.~\ref{Sect:corPBa} and \ref{Sect:fMpop}.


\section{Is there a correlation between barium intensity and orbital
period?}
\label{Sect:corPBa}

The existence of a correlation between the orbital separation (or more precisely, 
periastron
distance) and the level of
chemical  peculiarities would clearly be of key importance for understanding the
mass transfer process at work in the progenitor systems of barium stars. 
Unfortunately, these
quantities are not easily available for our complete sample, as not
all stars have been the target of detailed abundance analyses on one
hand, and on the other hand,  the orbital separations cannot be derived for
spectroscopic binaries with one observed spectrum without further assumptions.
Therefore, the ($P, \Delta(38-41)$) diagram presented in
Fig.~\ref{Fig:delta38P} has been used instead, since the $\Delta(38-41)$
color index is shown in the Appendix to provide a fairly good measure of the
heavy-element overabundances (see Fig.~A.1).
Although there is a general tendency for longer-period systems to exhibit less
severe peculiarities [i.e. larger $\Delta(38-41)$ values], as is
generally expected for the wind accretion process
(Sect.~\ref{Sect:fMpop} and Theuns et al. 1996), there is a
considerable scatter in the ($P, \Delta(38-41)$) diagram\footnote{The scatter is even more
severe if one
considers periastron distance instead of orbital period}.

\begin{figure}
\vspace{9cm}
  \vskip -0cm
  \begin{picture}(8,8.5)
    \epsfysize=9.8cm
    \epsfxsize=8.5cm
    \epsfbox{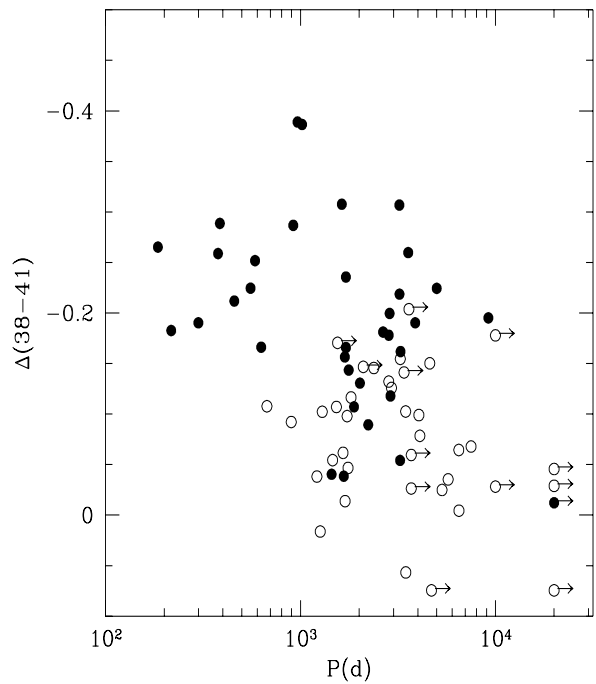}
  \end{picture}
  \vskip -0.5cm
\caption[]{\label{Fig:delta38P}
Orbital period vs $\Delta(38-41)$ index (see Appendix for details), for
mild and strong barium stars (represented by open and filled circles,
respectively). Arrows indicate that
only a lower limit is available for the orbital period. Stars
with only marginal evidence for binary motion 
(very low-amplitude variations, if any; Table~1b) have
been arbitrarily assigned a period of $10^4$~d, whereas constant stars have been
assigned a period of 2~10$^4$~d, for comparison 
}
\end{figure}

\begin{figure}
\vspace{9cm}
  \vskip -0cm
  \begin{picture}(8,8.5)
    \epsfysize=9.8cm
    \epsfxsize=8.5cm
    \epsfbox{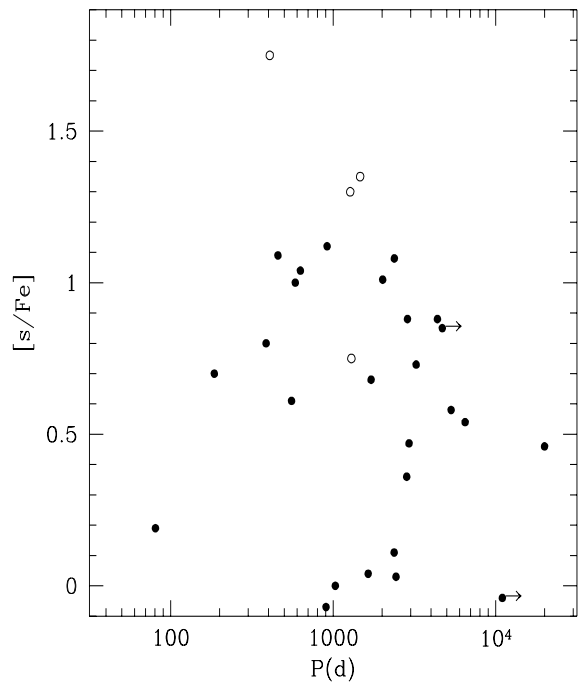}
  \end{picture}
  \vskip -0.5cm
\caption[]{\label{Fig:sFeP}
Orbital period vs heavy-element overabundance [s/Fe] (see
Table~\protect\ref{Tab:sFeP}). Filled dots correspond to barium and
normal giants in barium-like binary systems, whereas open dots
correspond to CH stars. 
Arrows correspond to lower limits on the orbital period. The constant-velocity
star HD 65854 has been arbitrarily assigned a period of 2~10$^4$~d 
}
\end{figure}

At any given orbital period, the scatter in $\Delta(38-41)$ is clearly larger than
would be expected solely from the scatter in the relation   
between $\Delta(38-41)$ and [YII/TiII]
(Fig.~A.1). The dispersion in Fig.~\ref{Fig:delta38P} is therefore  likely to
reflect real abundance variations at a given orbital period. To confirm that
suspicion, heavy-element abundances have been collected from the literature for
the barium stars of our sample, and are listed in Table~\ref{Tab:sFeP}.   
Figure~\ref{Fig:sFeP} presents the relation between the orbital period and [s/Fe],
where [s/Fe] = 0.5 ([Y/Fe] + [Nd/Fe]) whenever available.
It is unlikely that the large scatter observed in Fig.~\ref{Fig:sFeP} is due 
to systematic errors between abundances derived by different authors, 
as indicated by the comparison of 
HD~121447 ($P= 185$~d, [s/Fe] = 0.70, $\Delta (38-41) = -0.26$, K7Ba5) 
and HD~178717 ($P= 2866$~d, [s/Fe] = 0.88, $\Delta (38-41) = -0.20$, K4Ba5), 
both analyzed by Smith (1984).
Despite very different orbital
periods, the level of chemical peculiarities in these two stars 
is similar, as indicated not only by their heavy-element overabundances 
(Table~\ref{Tab:sFeP}) but also 
by their fluorine abundance (Jorissen et al. 1992a).
Moreover, several giants with {\it normal} heavy-element abundances, albeit  
in binary systems with barium-like orbital elements, are also shown on
Fig.~\ref{Fig:sFeP}. Their very existence
is another clear indication that the orbital separation 
is not the only parameter controlling the level of chemical peculiarities in
barium stars.
These barium-like binary systems involving a normal giant are listed   
in Table~\ref{Tab:sFeP} and are collected from 
the catalogue of Boffin et al. (1993), listing all 
spectroscopic binary systems with a giant component. That catalogue has been
searched for systems involving a red giant with 
normal heavy-element abundances according to McWilliam (1990), and falling in
the region $e < 0.1$, $P > 350$~d of the \elogP\ diagram (Fig.~\ref{Fig:elogP}). 
That region of the \elogP\ diagram is normally not populated by pre-mass-transfer
systems, as discussed in Sect.~\ref{Sect:elogP}, so that the unseen component in
these systems may be supposed to be a WD.  
No direct confirmation for the presence of a WD is however available for HD~33856
and HD~66216. For HD~13611 (=$\xi^1$~Ceti), B\"ohm-Vitense \& Johnson (1985)
reported some UV excess that they attribute to a WD companion, though that
statement has subsequently been questioned (Jorissen \& Boffin 1992; Fekel et
al. 1993). The clearest case of a post-mass-transfer system consisting of a WD and
a {\it normal} red giant is undoubtedly HD~160538 (=DR~Dra). Orbital elements ($P
= 904$~d, $e = 0.07$) have been obtained by Fekel et al. (1993), along with
definite evidence for the presence of a hot WD. An abundance analysis of that
star has recently been carried out by Berdyugina (1994), who concludes
that HD~160538 is a solar-metallicity giant ([Fe/H] $= -0.05$) with
normal Zr, Ba and La abundances. 

 
\begin{table*}
\caption[]{\label{Tab:sFeP}
Abundances of s-process elements in CH stars, barium stars and normal giant 
stars belonging to a 
barium-like binary system. The column labeled `elements'
lists the elements involved in [s/Fe]. Usually [s/Fe] = 0.5 ([Y/Fe] +
[Nd/Fe]); YII/TiII in that column means that [s/Fe] corresponds to
the [Y/Ti] ratio taken from McWilliam (1990; see Appendix for details).
The column labeled `Ref' provides the reference for the [s/Fe] data, 
according to the notes at the end of the table. The orbital periods of
CH systems are from McClure \& Woodsworth (1990)
}
\begin{tabular}{lllrrrlrllllllll}
\noalign{a. CH stars}\cr
\hline
HD & HR  & Sp. Typ. & $P$ (d) & $\Delta(38-41)$ & [s/Fe] & elements & Ref \cr
\hline\cr
209621 &  & CH       & 407     &        & 1.75 & Y, Nd & 7 \cr
224959 &  & CH       & 1273    &        & 1.30 & Y, Nd & 7 \cr
198269 &  & CH       & 1295    &        & 0.75 & Y, Nd & 7 \cr
201626 &  & CH       & 1465    &        & 1.35 & Y, Nd & 7 \cr
\hline\cr
\noalign{b. Barium stars}\cr
\hline\cr
77247   &     & G5 Ba1 & 80    & -     & 0.19 & Y, Nd & 3 \cr
121447  &     & K7 Ba5 & 185   &-0.26  & 0.70 & Y, Nd & 1 \cr
CpD $-64^\circ4333$&&K0 Ba4 & 386   &-0.29  & 0.80 & Y     & 1 \cr
46407   &2392 & K0 Ba3 &  457  &-0.21  & 1.09 & Y, Nd & 2 \cr
100503  &     & K3 Ba5 & 554   &-0.25  & 0.61 & Y     & 1 \cr
199939  &     & K0 Ba4 & 585   &-0.25  & 1.03 & Y, Nd & 3 \cr
44896   &     & K3 Ba5 &  629  &-0.17  & 1.04 & Y, Nd & 1 \cr
92626   &     & K0 Ba5 & 918   &-0.29  & 1.12 &Zr, Nd & 2 \cr
101013  &4474 & K0 Ba5 & 1711  &-0.17  & 0.68 & Y, Nd & 3 \cr
16458   & 774 & K1 Ba5 & 2018  &-0.13  & 1.01 & Y, Nd & 1 \cr
204075  &8204 & G5 Ba2 & 2378  &-0.15  & 1.08 & YII/TiII&5\cr
205011  &     & K1 Ba1 & 2837  &-0.13  & 0.36 & Y, Nd & 3 \cr
178717  &     & K4 Ba5 &2866   &-0.20  & 0.88 & Y, Nd & 1 \cr
131670  &     & K1 Ba1 &2930   &-0.13  & 0.47 & Y, Nd & 3 \cr
60197   &     & K3 Ba5 &3243   &-0.05  & 0.73 & Y, Nd & 1 \cr
196673  &     & K2 Ba2 &6500   & 0.00  &-0.19 &Zr, Nd & 3 \cr
199394  &     & G8 Ba1 &4606   &-0.15  & 0.88 & Y, Nd & 3 \cr
104979  &4608 & K0 Ba1 &$>4700$& 0.07  & 0.85 & Y, Nd & 4 \cr
139195  &5802 & K1 Ba1 &5324   &-0.02  & 0.58 &YII/TiII&5 \cr 
202109  &8115 & G8 Ba$<1$& 6489&-0.06  & 0.54 & YII/TiII&5\cr
98839   &4392 & G7 Ba$<1$&$>11000$& -  &-0.04 & Y, Nd & 8 \cr
65854   &     & K1 Ba3 & cst   & -     & 0.46 & Y, Nd & 3 \cr
\hline\cr
\noalign{c. Normal giants in barium-like binary systems}\cr
\hline
160538  &      & K0III + WD&  904 & -     &-0.07 &Zr, Ba, La & 6 & DR Dra\cr
33856   & 1698 & K0.5III   & 1031 & -     & 0.00 & YII/TiII& 5 \cr
13611   &  649 & G8III + WD?&1642 & 0.07  & 0.04 & YII/TiII& 5 &
$\xi^1$ Cet\cr                         
169156  & 6884 & K0III     & 2374 & -     & 0.11 & YII/TiII&5\cr
66216   & 3149 & K2III     & 2438 & -     & 0.03 & YII/TiII& 5 \cr
\hline\cr
\end{tabular}

References:\\
1: Smith 1984; 2: Kov\'acs 1985; 3: Za\v{c}s 1994; 4: Tomkin \& Lambert 1986;
5: McWilliam 1990; 6: Berdyugina 1994; 7: Vanture (1992); 8: Sneden et al. 
(1981).

\end{table*}

These counter-examples (especially DR~Dra) strongly suggest that {\it binarity
is not a sufficient condition to produce a barium star!} Another
condition -- like a low metallicity? -- 
seems thus required to form a barium star, and such a hidden
parameter may, at least in part, be responsible for the blurred
correlation between the orbital
period and the intensity of the chemical anomalies. Unfortunately, it is difficult
to evaluate the exact influence of metallicity on that correlation, because an
homogeneous set of metallicity determinations is lacking for the stars in our
sample. Large systematic
differences may indeed be present between the metallicity determinations by
different authors, as discussed e.g. by Busso et al. (1995). As an illustration,
metallicities ranging from $-$0.45 to +0.13 have been published for HD 46407! 
The location of CH stars in Fig.~\ref{Fig:sFeP} nevertheless lends strong support
to the idea that metallicity is the hidden parameter controlling the level of
chemical peculiarities at a given orbital period, since these low-metallicity
stars have the largest [s/Fe] values at any given orbital period. 
Kov\'acs (1985) also noticed that there is a correlation between
[Ba/Fe] and metallicity [Fe/H]: strong barium stars generally have a metallicity 
lower than mild barium stars 
(see also the recent compilation of Busso et al. 1995, and North et al. 1994 for
a similar finding among barium {\it dwarfs}). Clayton (1988) provided a 
theoretical foundation for that empirical finding: if 
$\an{13}{C}{16}{O}$ is the 
neutron source responsible for the operation of the s-process, its efficiency 
in terms of neutron exposure
is expected to be larger in a low-metallicity environment. Therefore, 
barium stars would be easier to produce in a low-metallicity
population. That question will be discussed in more details
in Sect.~\ref{Sect:fMpop}, since there are obviously other parameters involved (like the
dilution factor
and the amount of matter transferred from the former AGB companion) whose impact should be
evaluated as well. 

\section{The mass-function distributions}
\label{Sect:fM}
\subsection{Barium and CH stars}
\label{Sect:fMBa}

The cumulative frequency distribution of mass functions $f(M)$
conveys information on the masses of the  two components in the binary system,
since
\begin{equation}
f(M) =  \frac{M_2^3}{(M_1 + M_2)^2} \;\; \sin^3 i \equiv Q \sin^3 i,
\end{equation}
$M_1$ and $M_2$ being the masses of the red giant and of its
companion, respectively.
As shown by Webbink (1986) and McClure \& Woodsworth (1990), the $f(M)$
distribution of peculiar red giants like barium and CH stars is very different
from that of normal red giants.
The PRG distributions are in fact compatible with a narrow range of
$Q$ values, convolved with random orbital inclinations.
McClure \& Woodsworth (1990) obtained $Q = 0.041 \pm 0.010$~\Msun\ for
their sample of 16 barium stars, and $Q = 0.095 \pm 0.015$~\Msun\ for
CH stars.
Such narrow $Q$ distributions are indeed expected if the companions of
PRG stars
are WD stars with masses spanning a narrow range, as is the case for
field WDs (Reid
1996). On the contrary, since the companions of normal giants in
binary systems 
ought not be WDs, their masses may span a much wider range (the only constraint
being then
$M_2 \le M_1$), thus contrasting with the PRG $f(M)$ distributions.     

\begin{figure}
\vspace{9cm}
  \vskip -0cm
  \begin{picture}(8,8.5)
    \epsfysize=9.8cm
    \epsfxsize=8.5cm
    \epsfbox{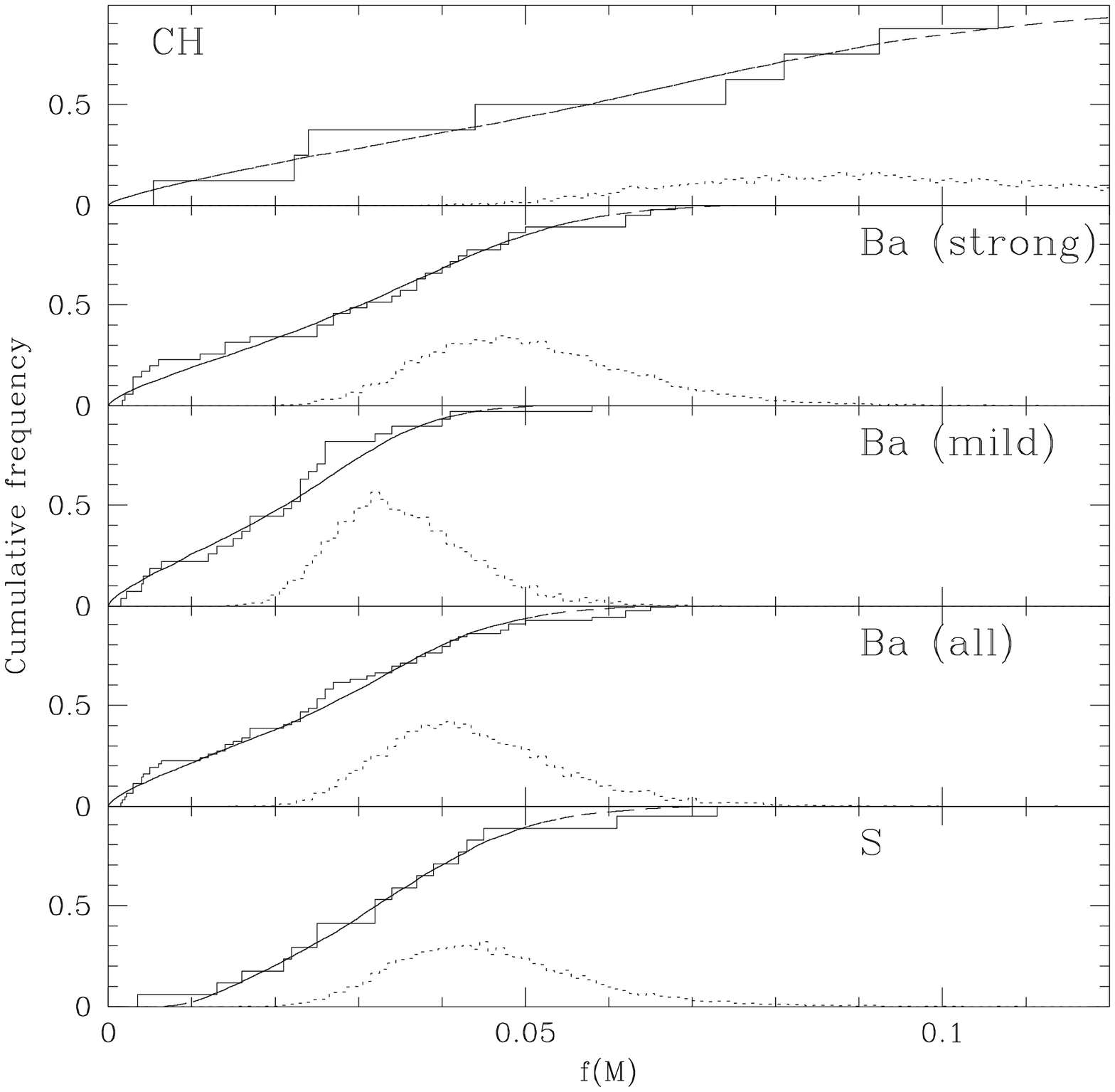}
  \end{picture}
  \vskip -0.5cm
\caption[]{\label{Fig:fMCHBa}
Comparison of the synthetic (dashed line) and observed (solid line) 
mass-function distributions
for CH stars (top panel, with data from McClure \& Woodsworth 1990), 
strong barium stars (second panel from top, from
Table~2a), mild
barium stars (third panel from top, from
Table~1a), all barium stars (fourth panel
from top), and S stars (bottom panel, from 
Table~3a). For S stars, the synthetic
distribution has been constructed by adopting a detection threshold $i >
36^\circ$ to simulate the deficit of systems with small mass functions
(see text).
The dotted line corresponds to the distribution of $Q$
}
\end{figure}

The larger samples considered
in this paper make it possible to derive the distributions of
$Q$ separately for mild and strong barium stars.    
These distributions have been extracted from the observed
distribution of $f(M)$ by two different methods. 
In the first method, 
the observed $f(M)$ distribution is simply compared to simulated
distributions assuming random orbital inclinations and gaussian
distributions (of mean $\overline{M}_i$ and standard deviation
$\sigma_i$, $i = 1,2$) for the masses $M_i$ of the two components.
Since $f(M)$ depends upon the masses only through the ratio $Q =
M_2^3/(M_1+M_2)^2$, 
almost equally good fits 
(expressed in terms of the greatest distance $D$
between the synthetic and observed $f(M)$ distributions) 
are obtained for  different
combinations of $\overline{M}_1$ and $\overline{M}_2$, all corresponding to
the same value of $\overline{Q}$. The value of $\overline{Q}$ minimizing $D$
for the different PRG families is listed in
Table~\ref{Tab:fM}. The synthetic and observed $f(M)$ distributions
are compared in Fig.~\ref{Fig:fMCHBa}. The best fits are obtained with
$\sigma_1 = 0.2$~\Msun\  and $\sigma_2 = 0.04$~\Msun.  

\begin{table}
\caption[]{\label{Tab:fM}
Average masses $\overline{M}_1$ of the giant star in various PRG families as
derived
from the cumulative frequency distribution of the mass functions,
for two different values of the companion average mass $\overline{M}_2$. $N$ is
the
number of available orbits
}
\begin{tabular}{llllllll}
Family & $N$ & $\overline{Q}$ & \multicolumn{2}{c}{$\overline{M}_1$} & Ref\cr
\cline{4-5}\smallskip\cr
 &&               & $\overline{M}_2 = 0.60$ & $\overline{M}_2
= 0.67$ \cr
            &&  (\Msun)      & (\Msun)  & (\Msun)\cr
\hline
\medskip\cr
CH              &  8     & 0.095  & 0.9 & 1.1 & 2 \cr
Barium (strong) & 36     & 0.049  & 1.5 & 1.9 & 1 \cr
Barium (mild)   & 27     & 0.035  & 1.9 & 2.3 & 1 \cr
Barium (total)  & 63     & 0.043  & 1.65& 2.0 & 1 \cr 
S (extrinsic)   & 17$^a$ & 0.041  & 1.6 & 2.0 & 1 \cr
\hline\cr
\end{tabular}

Remark: a: the two peculiar S stars HD~191589 and HDE~332077 were not
included (see Sect.~\ref{Sect:fMS})\\
References: 1. This work; 2: McClure \& Woodsworth (1990)
\end{table}

\begin{figure}
\vspace{9cm}
  \vskip -0cm
  \begin{picture}(8,8.5)
    \epsfysize=9.8cm
    \epsfxsize=8.5cm
    \epsfbox{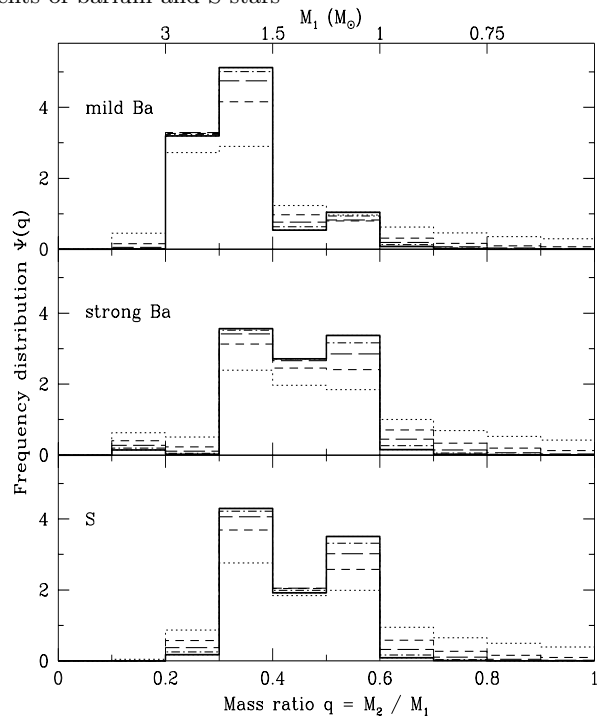}
  \end{picture}
  \vskip -0.5cm
\caption[]{\label{Fig:Lucy}
Distributions of mass ratios $q = M_2/M_1$ extracted from the observed
$f(M)$ distributions using the Richardson-Lucy iterative algorithm,
for mild barium stars (upper panel), strong barium stars (middle
panel) and S  stars (lower panel). During
the inversion process, $M_2$ has been fixed at 0.6~\Msun. The
corresponding mass $M_1$ of the giant star can be read from the upper
scale. The various lines refer to the distributions obtained after 2
(dotted line),
4 (short-dashed line), 6 (long-dashed line), 8 (dash-dotted line) and
10 (thick line) iterations 
}
\end{figure}

A similar analysis performed by North et al. (1997) on a sample of
{\it dwarf} barium stars provided the first independent estimate of
the average companion mass. Unlike the
barium giants, the barium dwarfs offer the possibility to directly
derive their masses
by fitting evolutionary tracks to their observed surface
gravities and effective temperatures. The known distribution of
primary masses and the average ratio $Q$ derived from
the $f(M)$ distribution then yield  
an average mass $\overline{M}_2$  of $0.67 \pm 0.09$~\Msun\ for the
companion. Assuming that dwarf barium stars, giant barium stars and
extrinsic S stars represent
successive stages along the same evolutionary path
(Sects.~\ref{Sect:S} and \ref{Sect:related}), the  $\overline{M}_2$ value derived by
North et al. (1997) for the companions of dwarf barium stars may be 
adopted to derive 
the average mass $\overline{M}_1$ of the giant 
for the different
PRG families listed in Table~\ref{Tab:fM}. 
For comparison,
the average mass $\overline{M}_1$ corresponding to $\overline{M}_2 =
0.60$~\Msun\  has also been listed.

The second method, described by Cerf \& Boffin
(1994) and based on the Richardson-Lucy iterative inversion algorithm,
fully confirms the above results. Figure~\ref{Fig:Lucy} shows the
extracted distributions of mass ratios $q=M_2/M_1$ and masses $M_1$
when $M_2$ is fixed at 0.6~\Msun. 

The $f(M)$ distributions of CH stars, mild barium stars and strong
barium stars are clearly distinct, and this difference is reflected in
the corresponding average masses $\overline{M}_1$ listed in
Table~\ref{Tab:fM}.
A Kolmogorov-Smirnov test confirms the significance of this
difference, since the null hypothesis that the distributions of mild
and strong barium stars are extracted from the same parent population
may be rejected with a first kind risk of only 0.6\%.

Although the difference in the $Q$ distributions of mild barium stars, strong
barium stars and CH stars may equally well result from a difference in the giant or
companion masses, the different kinematical properties
reported for these families suggest a difference in the respective
masses of the {\it giant} star.
The kinematics of CH stars is typical of a halo population (McClure
1984ab and references therein), whereas barium stars belong to a disk
population. Moreover, there are several pieces of evidence that
mild barium stars belong to a somewhat younger population than strong
barium stars. Catchpole et al. (1977; see also L\"u 1991) showed that 
the velocity dispersion of mild barium stars is smaller than that of strong
barium stars. From a statistical analysis of the positions of barium
stars in the Hertzsprung-Russell diagram based on Hipparcos
parallaxes, Mennessier et al. (1997) conclude that mild barium stars
are mostly clump giants with a mass in the range 2.5 -- 4.5 \Msun,
whereas strong barium stars populate the giant branch and have masses in the
range 1 -- 3 \Msun. These mass estimates are consistent with those
derived from the mass-function distributions (Fig.~\ref{Fig:Lucy}).

It may thus be concluded that mild barium stars, strong barium stars and
CH stars represent a sequence of increasingly older galactic populations.


\subsection{S stars}
\label{Sect:fMS}

\begin{figure}[t]
\vspace{10cm}
  \vskip -0cm
  \begin{picture}(8,8.5)
    \epsfysize=9.8cm
    \epsfxsize=8.5cm
    \epsfbox{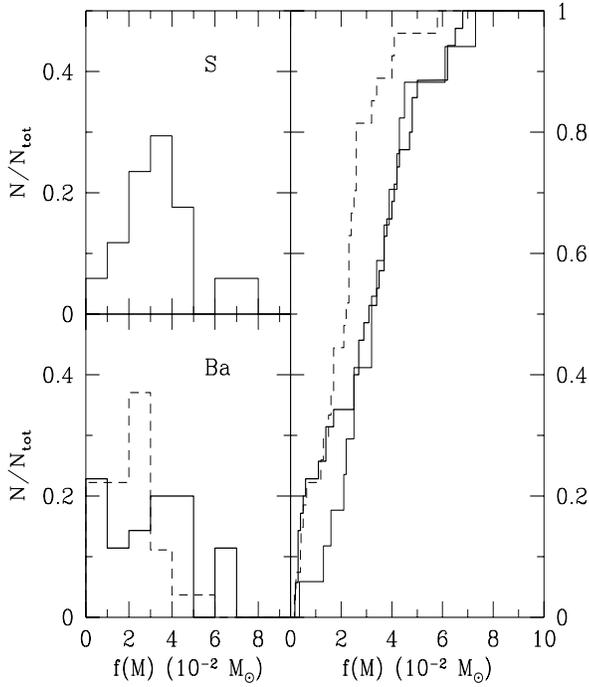}
  \end{picture}
  \vskip -0.5cm
\caption[]{\label{Fig:fMSBas}
Comparison of the  
mass-function distributions
of barium stars (thick solid line: strong barium stars; thin dashed
line: mild barium stars) and
S stars (thin solid line). 
The mass functions of HD 191589 and HDE 332077 were not included in
the comparison
}
\end{figure}

Figure~\ref{Fig:fMSBas} compares the cumulative distributions of mass
functions for barium and S stars. Apart from a deficit of systems
with small mass functions, the $f(M)$ distribution of S stars is
very similar to that of strong barium stars. 
The deficit of S systems with small mass functions 
$f(M)$ can probably be attributed  
to the difficulty of detecting small-amplitude binaries
for these stars with a substantial radial-velocity jitter
(Sects.~\ref{Sect:jitter} and \ref{Sect:uniform}, and
Figs.~\ref{Fig:Sbjitter} and \ref{Fig:isoproba}).  

Note that the two S stars HD~191589 and HDE~332077 were not included
in the comparison, since they have
very discrepant mass functions of 0.395 and 1.25~\Msun, respectively (Table~3a).
A-type companions were detected for these stars with the
{\it International Ultraviolet Explorer} (Ake \& Johnson 1992; Ake et
al. 1992), consistent with their mass functions.
From the current radial-velocity data, 
there is no indication that these systems might  be
triple.   The evolutionary status of these Tc-poor S stars is
currently unknown. 

\section{Mild vs. strong barium stars: entangled effects of 
orbital dynamics and galactic population}
\label{Sect:fMpop}
  
In the previous sections, mild and strong barium stars have been seen
to differ in many respects: (i) short-period ($P \la 1500$~d), nearly-circular
systems are lacking among mild barium stars, (ii) there is a tendency
for systems with longer periods to have milder chemical anomalies,
although the large scatter in that relation suggests that some other 
parameter (like metallicity?) partially controls the level of chemical
anomalies, (iii)
strong barium stars generally have lower metallicities than mild
barium stars, (iv) mild barium stars have smaller mass functions on
average, and (v) mild barium stars are kinematically younger than strong
barium stars. 

The previous facts consistently suggest that mild barium stars belong
to a younger, more metal-rich and more massive population than strong
barium stars. In this section, we investigate the consequences of this  
difference in galactic population on the mass-transfer
process. For  that purpose, the chemical anomalies  of mild and
strong barium stars are computed using a simple
dynamical model, assuming that the former AGB progenitor of the
present WD has dumped heavy-element-rich matter onto its companion. 
    
The parameters controlling the intensity of chemical anomalies in
barium stars may be easily identified with the aid of the
following formula, relating  
the overabundance factor $s$ of heavy elements in
the envelope of the barium star 
(i.e. the ratio between the
abundances after completion of the accretion and dilution processes,
and the initial envelope abundances) 
to their overabundance
$f$ in the material accreted from the former AGB star: 
\begin{equation}
\label{Eq:fg}
s = \frac{f \Delta M_{\rm Ba} + M_{\rm Ba,0} - M_{\rm Ba, core}}
         {  \Delta M_{\rm Ba} + M_{\rm Ba,0} - M_{\rm Ba, core}}
  \equiv f{\tens F} + ( 1 - {\tens F}),
\end{equation}  
where $\tens{F}$ is the dilution factor of the accreted matter 
$\Delta M_{\rm Ba}$ in the envelope of mass $(M_{\rm Ba,0} - M_{\rm Ba,
core} + \Delta M_{\rm Ba})$. 
Here, $M_{\rm Ba,0}$ and $M_{\rm Ba, core}$ denote the
initial total mass 
and core mass, respectively, of the accreting barium star.  
The accreted mass $\Delta M_{\rm Ba}$ is computed from the
relation 
\begin{equation}
\Delta M_{\rm Ba} = \beta (M_{\rm AGB,0} - M_{\rm WD}), 
\end{equation}
where $\beta$ corresponds to the fraction of the mass lost by the former AGB
star of mass $M_{\rm AGB,0}$ (at the start of the thermally-pulsing phase) 
that is actually accreted by the barium
star. For the sake of simplicity, it is assumed that all cases
considered below undergo wind accretion (so that predictions will be restricted to systems
with $P \ga 1500$~d, in accordance with the discussion of Sect.~\ref{Sect:Pdist}).
The Bondi \& Hoyle (1944) prescription will therefore
be adopted for computing the accretion efficiency $\beta$ 
(see Theuns et al. 1996 for more details):
\begin{equation}
\label{Eq:beta1}
\beta = \frac{\alpha}{A^2} \left( \frac{G M_{\rm Ba}}{v_{\rm w}^2}
                           \right)^2
        \frac{1}{[1+(v_{\rm orb}/v_{\rm w})^2]^{3/2}},
\end{equation}
or using Kepler's third law:
\begin{equation}
\label{Eq:beta2}
\beta = \alpha \mu^2 \frac{k^4}{[1+ k^2]^{1.5}},
\end{equation}
where $k \equiv v_{\rm orb}/v_{\rm w}$, $v_{\rm orb}$ and $v_{\rm w}$
being the relative orbital velocity ($2\pi A/P$)
and the wind velocity, respectively, $G$ is the gravitational
constant, $A$ is the orbital separation, $\mu \equiv  M_{\rm
Ba}/(M_{\rm AGB} + M_{\rm Ba})$, and $\alpha$ is a scaling
parameter. In the above relation, it
has been assumed that the wind is highly supersonic.   
Detailed hydrodynamic simulations (Theuns et al. 1996)
have shown that $\alpha \sim 0.05$ in the situation of interest here.    

The above relations thus indicate that the following parameters will have
some impact on the level of chemical anomalies:\\
(i) a large orbital separation will result in a small accretion cross
section (Eq.~\ref{Eq:beta1}, valid for a wind-accretion scenario). 
Note, however, that the functional dependence of $\beta$ with $A$ is more
complicated than the simple explicit $A^{-2}$ factor appearing
in Eq.~(\ref{Eq:beta1}), since there is an implicit 
dependency through the orbital velocity $v_{\rm orb}$;\\
(ii) a lower mass for the barium star results in a smaller dilution
(i.e. larger $\tens{F}$). However, the effect is opposite on the
accretion cross section (Eq.~\ref{Eq:beta1}), notwithstanding the
implicit dependence through $v_{\rm orb}$;\\
(iii) a smaller metallicity probably results in larger heavy-element
overabundances $f$ in the AGB progenitor (see the discussion at the
end of Sect.~\ref{Sect:corPBa}). As low-metallicity giants are expected to
have low masses, this effect is strongly coupled with (ii);\\
(iv) a larger mass for the AGB progenitor results in more mass being
lost and thus accreted by the companion; a more massive WD is also produced.

The above qualitative discussion shows that the various
relevant parameters are strongly coupled with each other, 
thus calling for detailed calculations. Because in the
framework of the simple wind-accretion model adopted here, the
accretion rate depends upon the orbital separation (see
Eq.~\ref{Eq:beta1}), the resulting overabundance $s$ must be computed
by taking into account the
variation of the orbital separation in the course of the mass transfer
process. Neglecting the anisotropy in the mass loss process induced by
the accretion, as well as a possible transfer 
of linear momentum from the wind to the accreting star (see Theuns et
al. 1996), the variation
of the orbital separation obeys the equation:
\begin{equation}
\label{Eq:dA}
\frac{\dot{A}}{A} = -\frac{\dot{M}_1+\dot{M}_2}{M_1+M_2}.
\end{equation}
The amount of accreted matter has been computed by integrating
Eqs.(\ref{Eq:fg})--(\ref{Eq:dA}) using a Runge-Kutta scheme, 
starting with $M_{\rm AGB,0}$ equal to $M_{\rm Ba}$ (the present  mass
of the barium star), and integrating till $M_{\rm AGB} = M_{\rm WD} =
0.67$~\Msun\ (see Table~\ref{Tab:fM}). 
This initial condition ensures that the AGB
star was initially slightly more massive than the barium star, and
evolved faster. The amount of mass accreted thus represents a lower
limit, as the AGB star might have been more massive initially.

\begin{figure}
\vspace{10cm}
  \vskip -0cm
  \begin{picture}(8,8.5)
    \epsfysize=9.8cm
    \epsfxsize=8.5cm
    \epsfbox{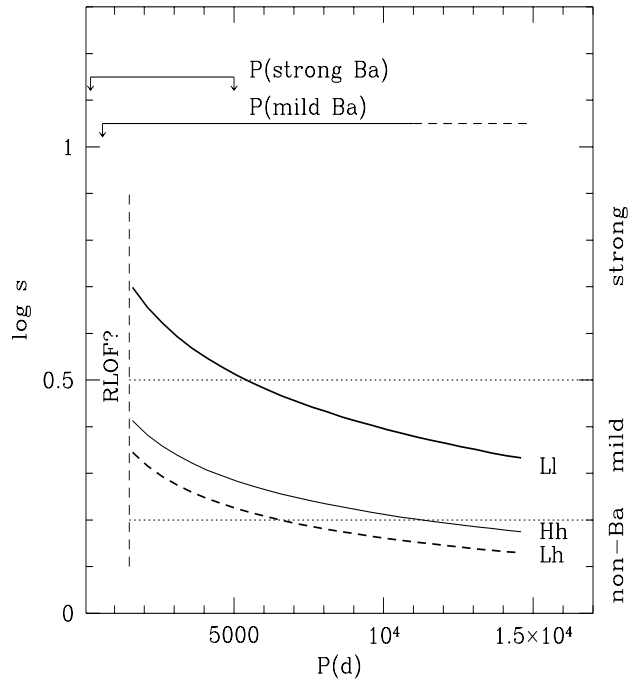}
  \end{picture}
  \vskip -0.5cm
\caption[]{\label{Fig:mildstrong}
Schematic predictions of the wind-accretion scenario
[Eqs.~(\protect\ref{Eq:fg}) -- (\protect\ref{Eq:dA})] for the cases
{\sc L}l, {\sc L}h and {\sc H}h defined in the text. The dotted lines
mark the (somewhat arbitrary) separation between strong, mild and
non-barium stars. The range in periods spanned by mild and strong
barium stars is represented by the horizontal lines in the upper part
of the figure
}
\end{figure}

Different combinations of paramaters have been considered, so as to
evaluate the relative importance of items (i), (ii) and (iii) listed above.
To evaluate the impact of the population difference between strong and
mild barium stars (see Table~\ref{Tab:fM}), cases with   
$M_{\rm Ba} = 1.8$ (mimicking strong barium stars, and labelled {\sc L} in
the following) and 2.4~\Msun\ (mimicking mild barium stars, and
labelled {\sc H} in the following) have been considered. 
To evaluate the impact of
metallicity (item iii), surface s-process overabundances $f$ of 130 and
40 in low-
and high-metallicity AGB stars, respectively, were considered 
(labelled $l$ and $h$, respectively, in the following).
The overabundance $s$ of heavy elements for cases {\sc H}h, {\sc L}h and
{\sc L}l is shown on
Fig.~\ref{Fig:mildstrong} as a function of the {\it final} orbital period $P$ (for $P \ga
1500$~d, as shorter final periods most probably involve RLOF, not adequately described by the
present scheme).
The overabundance $f$ of heavy elements in the accreted matter is
clearly of utmost importance in controlling the pollution level $s$.
In particular, it is clear that  different giant masses 
(Table~\ref{Tab:fM}) cannot explain the different pollution levels of
mild and strong barium stars (compare cases {\sc L}h and {\sc H}h on
Fig.~\ref{Fig:mildstrong}). 
The respective overabundances $s$ obtained in cases {\sc H}h and
{\sc L}h are even opposite to what is expected from the
observations, since the more massive mild barium stars ({\sc H}h) 
would be {\it more
polluted} than the strong barium stars ({\sc L}h) for a given $f$!
This contradiction results
from the fact that more mass was allowed to be exchanged in the
`mild barium' systems because the AGB progenitor was supposed to be
more massive in
the more massive mild barium systems than in the strong barium systems
(see above).

The mass of the AGB progenitor sensitively controls both the
amount of mass that 
can be exchanged (thus fixing the pollution level) and the mass
of the WD (thus fixing $Q$). It offers therefore a way to link the 
differences in pollution levels and mass functions that are observed in
mild and strong barium stars. However, because that explanation cannot
account for the kinematic differences observed between mild and strong barium
stars (Sect.~\ref{Sect:fMBa}), we favour an explanation in terms of a
difference  in the
$f$ values characterizing the matter accreted by the barium stars
belonging to the {\sc H} and {\sc L} populations having different metallicities. 
We find that, with the choice $f = 40$ and 130, 
the difference between mild and strong barium
stars is largely one of population, strong barium stars
belonging almost exclusively to a low-mass low-metallicity population
characterized by a large value of $f$
(represented by case {\sc L}l).  
Some merging between the {\sc H} and {\sc L} populations inevitably occurs
for mild barium stars, however, since they contain all stars with 
$0.2 \la \log s \la 0.5$ (see Fig.~\ref{Fig:mildstrong}). 
This prediction receives some support from the mass distribution 
of mild barium stars presented in Fig.~\ref{Fig:Lucy}. Although mild
barium stars are dominated by high-mass objects ($M_1 \sim 1.5$ -- 3
\Msun, corresponding to the
{\sc H} population in the above simulations), there is a small tail of
less massive objects which suggests that mild barium stars
are indeed a mixture of populations {\sc H} and {\sc L}.

The above values of $f$ appear plausible in view of the s-process
overabundances reported by Utsumi (1985) and Kipper et al. (1996) 
in N-type carbon stars, supposed to be
representative of the former AGB mass-losing star in the barium
systems. However, the parameter $f$ is admittedly not very tightly
constrained by our simple model. It is therefore encouraging that the value
of $f$ chosen to reproduce the upper period cutoff of
strong barium stars (in case {\sc L}l) predicts that mild barium stars 
can be produced in systems with periods up to a few $10^4$ d which is
consistent with the longest periods observed among mild barium stars
(Sect.~\ref{Sect:MCdiscussion}). 

Another argument related to CH stars lends strong support to the
idea that $f$ increases in low-metallicity populations, and 
is the principal factor controlling the pollution levels of PRG stars.  
Although the periods of CH stars span the same range as those of
barium stars (Fig.~\ref{Fig:elogP}), their heavy-element
overabundances at a given period are generally larger than those of
barium stars (Fig.~\ref{Fig:sFeP} and Table~\ref{Tab:sFeP}). 
Applying the method described above to CH stars, it is found that
values of $f$ of the order of 300 at least 
are required to account for the observed overabundance levels $s$.
The operation of the s-process must therefore have been very efficient
in the low-metallicity AGB star responsible for the pollution of 
the CH star envelope (see also the discussion in
Sect.~\ref{Sect:corPBa}). There is a very distinct property of the
heavy-element abundance distribution observed in CH stars that
confirms this fact. Heavy s-process
elements (`hs') like Ba are comparatively more overabundant than light
s-process elements (`ls') like Y, with [hs/ls]$ \ge 0.4$ (Vanture
1992). This situation is only encountered when the neutron exposure $\tau$
which characterizes the efficiency of the s-process reaches 
values as large as 1 mb$^{-1}$ (Vanture 1992). Such values lead to 
overabundances of heavy s-elements that are comparatively larger than   
for smaller $\tau$ (e.g. Fig.~5 of Busso et al. 1995), 
in agreement with the above requirement for large $f$ values in CH systems. 

\section{Short-period Ba systems and the question of RLOF}
\label{Sect:RLOF}

The mass-transfer model [Eqs.~(\ref{Eq:fg}) --
(\ref{Eq:dA})] used in Sect.~\ref{Sect:fMpop} to evaluate the
respective importance of the various parameters entering the problem 
is extremely schematic. In particular, it assumes that mass
transfer occurs through wind accretion. 
That assumption is clearly not valid at the short end of the period
distribution, where 
Roche lobe overflow (RLOF) enters the stage 
(see Sect.~\ref{Sect:Pdist}; Han et al. 1995; Jorissen et al. 1995). 
In fact, the wind-accretion scenario considered in
Sect.~\ref{Sect:fMpop}
 predicts that the system widens as it loses mass. 
For systems with periods shorter than $\sim 1500$~d, that
prediction is clearly incompatible with the requirement that the system
remained detached during mass transfer (see also the discussion in
Jorissen et al. 1995). Adopting for instance $M_{\rm AGB} = 0.6$~\Msun,
$M_{\rm Ba} = 1.5$~\Msun\ and $P = 1500$~d yields a Roche radius of
about 200~\Rsun, much smaller than typical radii of $\sim
500$~\Rsun\ observed for cool (\Teff $\sim 2200$~K) Mira stars (e.g. van Belle
et al. 1996). In such circumstances, it seems unavoidable that at
least the shortest-period systems among barium stars result from RLOF.
That conclusion is puzzling, however, in view of
the catastrophic outcome generally associated with RLOF when the
mass-losing star is the more massive  one and has a deep convective
envelope as is the case for AGB stars. Mass transfer then occurs on
a dynamical time scale (`unstable case C' mass transfer), and leads,
via a common envelope stage, to
the formation of cataclysmic variable stars with orbital periods of a
few hours (e.g. Meyer \& Meyer-Hofmeister 1979; de Kool 1992; Iben \&
Tutukov 1993; Ritter 1996). Barium stars must somehow have avoided this
evolutionary channel. In the remainder of this section, several ways
out of this channel are briefly suggested.
 
First, it is possible that the common envelope phase developing during
unstable case C mass transfer does not lead to a dramatic orbital
shrinkage if mechanisms internal to the AGB star reduce the effective binding
energy of its envelope (like the recombination energy in the hydrogen and helium
ionization zones, excitation of non-radial pulsation modes,
shock-heating, dust-driven winds...; Iben \& Livio 1993).
In those cases, not so much energy ought to be extracted from the
orbit to expell the common envelope, thus reducing (and even perhaps
suppressing) the orbital decay.
Due consideration of the thermal energy of the AGB envelope reduces
its effective binding energy and 
the orbital shrinkage associated with a common-envelope phase is
therefore limited. 
Han et al. (1995) have shown that   
barium systems like HD 77247 with periods as short as 80~d may form
under such circumstances.

Second, the dynamical instability associated with case C mass transfer
is suppressed when the mass-losing star is
{\it less massive} than its companion (Pastetter \& Ritter 1989), 
with $q  < q_{\rm crit} < 1$, where $q
= M_{\rm AGB}/M_{\rm Ba}$ and $q_{\rm crit}$ is given by 
Hjellming \& Webbink (1987; see also Hjellming 1989).
This situation is encountered if a  strong mass loss by wind steadily
reducing the mass of the AGB
star reversed the mass ratio prior to the onset of RLOF.  
Tout \& Eggleton (1988) have shown that the mass ratio is easily
reversed if the mass loss rate of the AGB star is tidally
enhanced by the companion (`Companion-Reinforced Attrition Process',
CRAP) so to exceed 
the Reimers rate by one or two orders of magnitude. 
By including this effect, Han et al. (1995)
were able to stabilize the RLOF of many systems, which ended up as
barium systems with periods in the range 250 -- 2500~d. 

Finally, the very concept of the Roche lobe may be irrelevant for
systems involving a mass-losing star where the wind-driving force may
substantially reduce the effective gravity of the mass-losing star (see the
discussion in Theuns \& Jorissen 1993).
The existence of an extra-force driving the mass loss
will clearly modify the shape of the equipotential surfaces, as shown by
Schuerman (1972). The geometry of X-ray 
binaries involving mass-losing supergiants (Bolton 1975; Kondo et
al. 1976), and possibly
also of the yellow symbiotic-barium star BD$-21^\circ3873$ (Smith et
al. 1997), are indeed observed to deviate from the predictions made
with the usual Jacobi
integral describing the restricted three-body equipotential surfaces. 
In particular, if the effective gravity is reduced
below some threshold depending upon the mass ratio, the critical Roche
equipotential will degenerate into a surface including {\it both} the
Lagrangian points $L_1$ and $L_2$ (if the
mass-losing star is the more massive component) or $L_1$ and $L_3$ 
(in the opposite case). Matter will thus not necessarily be confined
to the lobe surrounding the accreting component, but may escape from
the system through $L_2$ or $L_3$, after forming a circumbinary disk.  
Furthermore, in the case of a strong wind, the particles have
initially non-zero kinetic energy taken from
the internal energy of the mass-losing star, that makes the
equipotentials escapable barriers.
Therefore, only a limited fraction of the mass lost by the AGB star  
is accreted by the companion, and such a reduced accretion rate will
be less prone to trigger the expansion of the accreting star envelope,
at the origin of the common envelope formation.
The mass flows in these situations are complex, as shown by recent
{\it Smooth Particle Hydrodynamics} simulations (Theuns \& Jorissen
1993; Theuns et al. 1996).  

\section{Do S stars evolve from barium stars?}
\label{Sect:S}

The orbital properties of S stars appear to be identical to those of barium stars,
as apparent from Fig.~\ref{Fig:elogP} [$(e, \log P)$ diagram] and
Figs.~\ref{Fig:fMCHBa} and \ref{Fig:fMSBas} (mass-function distribution). 
This similarity is a strong indication that
binary S stars are simply the descendants of barium stars as they cool
while ascending either the RGB or the E-AGB\footnote{Strictly speaking, the S and
barium phases may even be intermingled in the sequence Ba (lower RGB) - S (upper
RGB) -- Ba (He clump) -- S (E-AGB and TP-AGB)}. Because the evolution
is slower on the RGB than on the E-AGB, it may actually
be expected that binary S stars be dominated by stars on the upper RGB rather than
on the E-AGB. In this case, {\it they ought to be low-mass stars} (because only
low-mass stars develop a red giant branch), in
agreement with the value 1.6~\Msun\ derived from their average $Q$
(Table~\ref{Tab:fM}).    

It is expected that at some point in their evolution on the RGB the
S stars with the shortest periods ($P \la 600$~d) will overflow their Roche lobe. 
In that respect, it should be noted that the hottest (i.e. least evolved) 
S star in our sample, HD 121447, has the shortest period among S
stars. The detailed analysis of this system performed by Jorissen et al. (1995)
concludes that the giant will overflow its Roche lobe before reaching the RGB-tip,
with a dramatic fate for the binary system as described in Sect.~\ref{Sect:RLOF} 
in relation with unstable case C mass transfer. The various processes that      
prevented the orbital decay of the binary system in a former stage of its
evolution, when 
the AGB progenitor of the present WD filled its Roche lobe, are no longer applicable
to this second RLOF event. First, the mass ratio can no longer be
inverted easily,
as the companion is already a low-mass WD, so that case C mass transfer is
necessarily dynamically unstable. Second, the envelope of a RGB star is more
tightly bound than that of an evolved AGB star (since it is less extended and its
`gravity-reducing' wind is much weaker), so that more energy has to be extracted
from the orbit to expell the common envelope, resulting in a stronger orbital
decay. A lack of short-period S systems is predicted from these considerations,
as already suspected by Jorissen \& Mayor (1992). 
This effect is not immediately apparent from the comparison between the cumulative
period distributions of barium and S stars (Fig.~\ref{Fig:PS}). However, there are
good reasons to suspect that the two shortest-period S systems, HD~121447 and
HD~191589, are peculiar in some respect (HD~121447: see above; HD 191589 is a
puzzling Tc-poor S star with an A-type companion, see
Sect.~\ref{Sect:fMS}). If these special cases are removed from the sample of S
stars, S systems with $P \la 600$~d appear to be lacking, in agreement with the
critical periods expected for RLOF on the upper RGB (see Table~5 of Jorissen \&
Mayor 1992).
 

\section{Related families: dwarf barium stars, post-AGB stars,
symbiotic stars}
\label{Sect:related}

Possible links between the PRG families considered in this paper and symbiotic
stars, post-AGB stars and dwarf barium stars are briefly considered in this
section.

\subsection{Dwarf barium stars}

Most barium stars may be expected to form as dwarf stars, since the stellar
lifetime is longer on the main sequence than on the giant branch, and since the
cross section for wind accretion is independent of the geometric radius of the
star according to the Bondi-Hoyle formula (Eq.~\ref{Eq:beta1}).  
Dwarf barium stars long remained elusive, until Luck \& Bond (1982, 1991) and
North et al. (1994) recognized that some of the CH subgiants previously identified
by Bond (1974), as well as some of the F dwarfs previously classified by Bidelman
(1985) as having a `strong Sr $\lambda4077$' line, had the proper abundance
anomalies, gravities and galactic frequencies to be identified with the long-
sought Ba dwarfs. A large fraction of binaries (about 90\%) has been found among
the stars with strong anomalies, as expected (McClure 1985; North et al. 1997).
The suspected WD companions to these dwarf barium stars appear to be too cool to
be detectable with the IUE satellite (Bond 1984; North \& Lanz 1991). 

On the contrary, it is the presence of a hot WD companion to the K dwarf
2RE J0357+283 that led
Jeffries \& Stevens (1996) to suspect it might be a dwarf barium star. A subsequent
spectroscopic analysis (Jeffries \& Smalley 1996) confirmed this early suspicion.
This star  is the first example of the new class of WIRRing (`Wind-Induced Rapidly
Rotating') stars, that have been spun up by wind accretion, as predicted by SPH
simulations (Theuns et al. 1996). The giant barium star HD~165141, which has a hot
WD companion as well (Fekel et al. 1993), is another member of that class
(Jorissen et al. 1996).    
 
\subsection{Post-AGB stars}

The post-AGB stars with large Fe deficiencies studied by Van Winckel
et al. (1995) are all binaries, and their $(e, \log P)$ diagram is very similar to
that of barium stars (Van Winckel et al. 1997). These post-AGB stars could
be related to barium stars, were it not for the absence of clear
evidence for s-process enrichment of these stars (Van Winckel 1995).

\subsection{Yellow symbiotic stars}

Symbiotic systems consist of a late-type giant and a hot companion, a WD in most
cases, exhibiting nebular lines in their spectra (e.g. Kenyon 1986). 
Yellow symbiotics involve a G or K giant and are often halo objects (Schmid \&
Nussbaumer 1993). Many exhibit the barium syndrome (UKS-Ce1 and S32: Schmid 1994;
AG Dra: Smith et al. 1996; BD$-21^\circ3873$: Smith et al. 1997), which must have
formed in much the same way as in barium and S systems.
These yellow symbiotics are probably evolving on the upper RGB or on the E-AGB,
thus being the Pop.II counterparts of the extrinsic S stars, whereas the hotter
CH stars would be the analogs of the barium stars.
A more detailed comparison can be found in Jorissen (1997).

\section{Conclusions}
\label{Sect:conclusion}

Radial-velocity monitoring of a complete sample of barium stars
with strong anomalies reveals that 35 out of 37 stars show clear
evidence of binary motion. For mild barium stars, that frequency
amounts to 34/40 (or 37/40 if one includes the suspected binaries).    
A Monte-Carlo simulation shows that these frequencies are compatible 
with the hypothesis that {\it all} the
observed stars are binary systems, some of them remaining undetected
because of unfavourable orbital orientation or time sampling. 
We conclude therefore that there is no need to invoke a barium-star
formation mechanism other than one (like mass transfer) directly related to the binary
nature of these stars. In other words, {\it binarity appears to be a
necessary condition to form a barium star}. It seems, however, that it
is not a sufficient condition, since binary systems with barium-like
orbital elements but no heavy-element overabundances seem to exist
(e.g. DR Dra). It has been argued that a metallicity lower than solar may be the
other parameter required to form a barium star. The increasing  levels of
heavy-element overabundances observed in the sequence mild Ba -- strong
Ba -- CH stars support that hypothesis, since this sequence is also one
of increasing age (and thus, to first order, of decreasing
metallicity), as revealed by their kinematical properties.

The $(e, \log P)$ diagram of PRG stars clearly shows the signature of mass
transfer, since the maximum
eccentricity observed at a given orbital period is much smaller than in a
comparison sample of normal giants in clusters.
Mass transfer rather than some
non-standard internal mixing induced by binarity must thus be held
responsible for the chemical peculiarities exhibited by PRG stars.  
A distinctive feature of the \elogP\ diagram of barium stars is the
presence of two distinct modes: a short-period mode ($P \la 1500$~d)
comprising nearly-circular orbits ($e < 0.08$), populated by strong
barium stars only, and  a long-period mode made exclusively of
non-circular orbits. At this point, it is not clear whether the
nearly circular orbits of the short-period mode bear the signature of
RLOF, or whether they arise from tidal circularization on the RGB
long after the mass-transfer process. Detailed models of binary
evolution are required to answer that question.  

The comparison of the mass-function distributions of mild and strong
barium stars confirms that  
the difference between them is mainly one of galactic
population rather than of orbital separation, since mild barium stars
host more massive giants than strong barium stars.
The loose correlation that is observed between the orbital period and
the level of heavy-element overabundances is another indication that a
parameter not directly related to the orbital dynamics has a strong
impact on the pollution level of the barium star. All these facts  
fit together nicely if the s-process operation is more efficient in a
low-metallicity population. 
Provided that the reaction $\rm ^{13}C(\alpha,n)^{16}O$ is the neutron
source responsible for the operation of the s-process, such a
correlation between metallicity and s-process efficiency is indeed
predicted on very general theoretical grounds. 
In this framework, the giant star in
Pop.II CH systems has accreted material much enriched in
heavy elements by its former AGB companion. Therefore, 
stars of old, low-metallicity populations like CH
stars (and to a lesser extent, strong barium stars) exhibit, {\it for a
given orbital period}, much larger
heavy-element overabundances than stars belonging to a younger population.

A radial-velocity monitoring of S stars confirms that 
Tc-poor S stars are all binaries and are the cool descendants of the
barium stars on the RGB or E-AGB, since
they have similar orbital periods and mass functions. There is a suggestion, however, that
the short-period tail of S stars may be truncated at about 600~d due to RLOF
occurring on the upper RGB. Similarly, for Pop.II stars, yellow
barium-symbiotic systems like AG Dra and BD$-21^\circ$3873 are the
cool descendants of the hotter CH stars.    
Two Tc-poor S stars, HDE 332077 and HD 191589, have unusually large
mass functions, and an A-type companion has been detected in both
cases with IUE, as well as in 57 Peg (Van Eck et al. 1998). 
The evolutionary status of these stars is
currently unclear. 

\acknowledgements 
This paper was written in part when A.J. was {\it Visiting Research Fellow}
at the Department of Astrophysical Sciences at Princeton University.
Financial
support from the {\it Fonds National de la Recherche Scientifique} 
(Belgium, Switzerland)
is gratefully acknowledged. S.V.E. is Boursier
F.R.I.A. (Belgium) and was partly
supported by a grant from the Soci\'et\'e Suisse d'Astronomie during a
stay at the Observatoire de Gen\`eve. The operation of the CORAVEL
spectrovelocimeter has been possible thanks to financial support from the {\it Fonds
National Suisse de la Recherche Scientifique}. 

\section*{Appendix: Setting up a photometric index probing the
chemical  peculiarities in barium stars}
\renewcommand\thefigure{A.\arabic{figure}}
\setcounter{figure}{0}

To be able to compare in a homogeneous way the level of chemical
peculiarities  in different barium stars is
clearly of paramount importance when trying to identify the
mass transfer scenario that operated in these stars. A key
constraint may indeed be obtained by assessing  whether or not
there is a correlation between the level
of chemical peculiarities and some dynamical parameter like  
the orbital period. As detailed abundance analyses are only
available for a small subset of barium stars, the aim of this Appendix
is to design 
a photometric index available for (nearly) all barium stars
and probing their chemical peculiarities. 

L\"u et al. (1983) and L\"u (1991) pointed out that barium
stars segregate according to their Ba index (as
defined in the scale of Warner 1965 from visual inspection of the
strength of the Ba$\lambda 4554$ line)  
in the $[C(38-41), C(42-45)]$ color-color diagram
constructed from DDO photometric indices (see McClure \& van den Bergh 1968).
This segregation is caused by the so-called Bond-Neff
depression, a broad absorption feature present in the
spectrum of barium stars and extending from about 350 to 450~nm (Bond \& Neff
1969; L\"u \& Sawyer 1979). A veil of many 
heavy-element lines has been proposed as the origin of this broad feature
(McWilliam \& Smith 1984; Wing 1985, and references therein), though
that issue is still controversial as CN bands are also important
contributors in this spectral region (Tripicco \& Bell 1991).    

\begin{figure}
\vspace{9cm}
  \vskip -0cm
  \begin{picture}(8,8.5)
    \epsfysize=9.8cm
    \epsfxsize=8.5cm
    \epsfbox{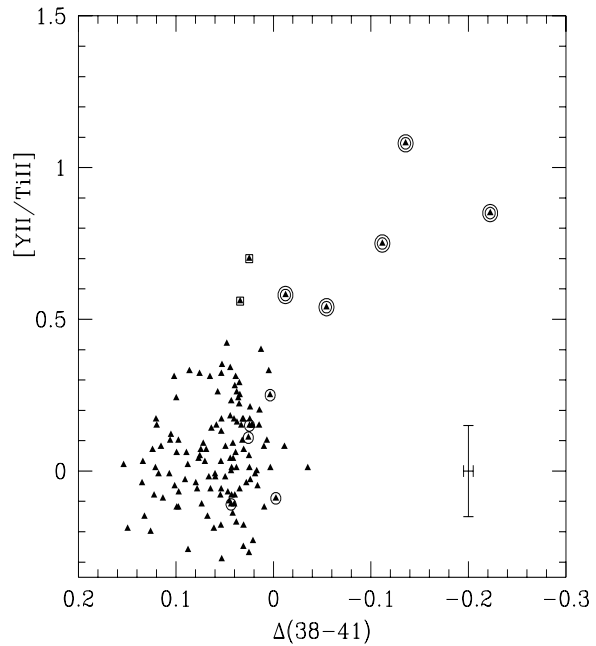}
  \end{picture}
  \vskip -0.5cm
\caption[]{\label{Fig:YTi}
The $\Delta(38-41)$ color index vs. [YII/TiII] for 131 G -- K giants with DDO
colors from McClure \& Forrester (1981) and abundances from McWilliam (1990).
Typical error bars have been indicated. 
Double-circled triangles identify the barium stars HD~116713,
HD~139195, HD~204075, HD~202109 and HD~212320, 
whereas squares correspond to heavy element-rich stars according to
McWilliam (1990), not previously reported as barium stars (see text).
Strong CN stars from the list of Keenan, Yorka \& Wilson (1987) have
been represented by single-circled triangles
}
\end{figure}

According to these findings, it may be expected that a photometric index of
the form $\Delta(38-41) = C(38-41) - m  C(42-45) + n$  
be related to the level of chemical peculiarities in barium stars
(see Fig.~4 of L\"u 1991).  
The $m$ and $n$ parameters in the above relation may actually be
chosen so as to yield the maximum correlation between $\Delta(38-41)$ and
some given abundance indicator. 
McWilliam (1988, 1990) showed that the abundance ratio YII/TiII,  derived
from the YII $\lambda 6795.4$ and TiII $\lambda 6607.0$ lines,
is a powerful indicator for detecting heavy-element abundance
peculiarities in red giants, since (i) that abundance ratio is
relatively insensitive to the atmospheric parameters, and (ii) the distribution
of YII/TiII ratios in a sample of about 600 G -- K giants is very
narrow, with a FWHM of only 0.3~dex centered at log(YII/TiII) $= -2.75$ 
(adopted as normalization value for [YII/TiII] in the following).

The $\Delta(38-41)$ index has therefore been calibrated in terms of 
[YII/TiII] abundances, 
using 131 G and K giants for which both DDO colors (McClure \&
Forrester 1981) 
and abundances (McWilliam 1990) are available.\\  


\begin{figure}
\vspace{9cm}
  \vskip -0cm
  \begin{picture}(8,8.5)
    \epsfysize=9.8cm
    \epsfxsize=8.5cm
    \epsfbox{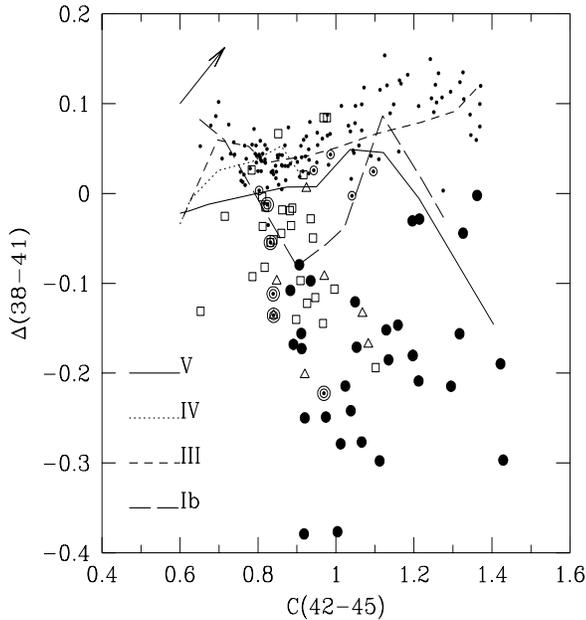}
  \end{picture}
  \vskip -0.5cm
\caption[]{\label{Fig:delta3841}
The $\Delta(38-41)$ color index vs. $C(42-45)$ for barium stars (open
squares: Ba0 -- Ba1, open triangles: Ba2 -- Ba3, filled circles: Ba4
-- Ba5). The loci of main sequence stars (class V), subgiant stars
(class IV), giant stars (class III) and supergiant stars (class Ib),
as provided by McClure \& Forrester (1981), are indicated by the
various lines. Small dots correspond to 131 G -- K giants common to
the samples of McClure \& Forrester (1981) and McWilliam (1990), as
plotted in Fig.~\protect\ref{Fig:YTi}. Double-circled and
single-circled symbols identify the same stars as in 
Fig~\protect\ref{Fig:YTi}. The arrow corresponds 
to a reddening by $E_{\rm B-V} = 0.5$ 
}
\end{figure}

The value $m=0.90$ is obtained by requiring maximum linear correlation
between [Y/Ti] and $\Delta(38-41)$ for that sample, and $n = 1.33$ 
ensures that barium stars have $\Delta(38-41) < 0$, whereas nearly all
normal giants have $\Delta(38-41) \ge 0$ (Fig.~\ref{Fig:YTi}).\\


Although the expected trend is clearly present (taking into account
the $\sim 0.2$~dex uncertainty on [Y/Ti]), there are a few stars
in `forbidden' regions [namely (i) non-barium stars [Y/Ti]$ <
0.5$ with $\Delta(38-41) < 0$, or (ii)
possible barium stars [Y/Ti]$ > 0.5$ with $\Delta(38-41) \ge 0$], 
degrading the ability of the $\Delta(38-41)$ index to identify barium stars. 
The two stars (HR~7754 and HR~8590) 
in region~(ii) were not previously identified as barium
stars. HR~7754 is in fact present in the list of MK standards
provided by Keenan \& McNeil (1989), and there is no mention
whatsoever of the barium nature of that star, classified as G9III. 
Apart from the fact that
HR~7754 is member of a complex multiple visual system, there is
currently no clue as to the origin of this discrepancy.        

Stars in region~(i) ([Y/Ti]$ < 0.5$, $\Delta(38-41) < 0$) 
may actually be bright giants or CN-strong stars,
spilling somewhat over into the region of barium stars.
It is indeed clear from the [$C(42-45), \Delta(38-41)$] diagram 
(Fig.~\ref{Fig:delta3841}), where the fiducial loci of dwarfs, giants and
Ib supergiants from McClure \& Forrester (1981) have been indicated,
that Ib supergiants with $\Delta(38-41) < -0.1$ may be found among
barium stars. As CN bands are strong contributors to  the
the $C(38-41)$ index (Tripicco \& Bell 1991),
CN-strong stars tend to have 
$\Delta(38-41)$ indices smaller than average as well, and some (like HR~3905) 
may also contaminate region (i) (Fig.~\ref{Fig:delta3841}).

No reddening correction has been applied to the stars plotted
in Fig.~\ref{Fig:delta3841}. According to the reddening correction
factors provided by McClure (1979), reddening is expected to have a
very limited impact on the $\Delta(38-41)$ index, as the de-reddened
index writes $\Delta_0(38-41) = \Delta(38-41) - 0.13 E_{\rm B-V}$.    

Despite ambiguities in identifying barium stars when
$\Delta(38-41)$ is close to 0, smaller values of that index correlate
fairly well with heavy-element overabundances, as indicated
by the barium stars present in Fig.~\ref{Fig:YTi}. In fact, a similar
correlation was already obtained for the closely-related $\Delta
c_1$  index defined by Jorissen et al. (1992b) from
Str\"omgren photometry. 

With the adopted normalization,
mild barium stars typically have $-0.1 \la \Delta(38-41) \la 0$
while strong barium stars (with Ba4 and Ba5 indexes)
have $\Delta(38-41) \la -0.1$.

\clearpage
\newpage


\def\baselinestretch{1}
\setcounter{table}{0}

\setlength\textheight{19cm}

\renewcommand{\thetable}{\arabic{table}a}
\begin{table*}
\caption[]{\label{Tab:orbiteBam}
Orbital elements for mild (Ba$<$1, Ba1 and Ba2) barium stars. 
Column 2 provides the spectral
subclass ($>0$ if K type, $<0$ if G type) and column 3 the Ba index,
from L\"u et al. (1983). 
The columns labeled  $\overline{\epsilon}_1$ and $N$ give the average error on one
measurement and the number of measurements, respectively.
A dash in column $Sb$ 
indicates that the spectral line width is smaller than the instrumental profile.
For orbits obtained from instruments other than CORAVEL, 
the $Sb$ parameter is not available (`na'). When an orbital solution is available,
$\gamma$ is the systemic radial velocity; otherwise, it is the average radial-velocity with its 
standard deviation.
$\Delta(38-41)$ is a photometric  index characterizing 
the strength of the Ba anomaly (see text). The numbers in column `Ref.'
refer to Table~5, which gives the reference where the complete set of
orbital parameters for the considered system may be found 
}
\tabcolsep 4pt
{\small
\begin{tabular}{lrll@{$\pm$}ll@{$\pm$}ll@{$\pm$}lccccr@{$\pm$}rcl}
\hline\cr
HD/DM & Sp. & Ba & 
\multicolumn{2}{c@{\mbox{}}}{$P$}      &
\multicolumn{2}{c@{\mbox{}}}{$e$}      & 
\multicolumn{2}{c@{\mbox{}}}{$f(M)$}   &
$O-C$ & $\overline{\epsilon}_1$ & $N$ & $S_b$ &
\multicolumn{2}{c@{\mbox{}}}{$\gamma$} & 
$\Delta(38-41)$ & Ref. \cr

 & & &
\multicolumn{2}{c@{\mbox{}}}{(d)}      &
\multicolumn{2}{c@{\mbox{}}}{}      & 
\multicolumn{2}{c@{\mbox{}}}{(\Msun)}   &
(km/s) & (km/s) & & (km/s) & 
\multicolumn{2}{c@{\mbox{}}}{(km/s)} &  &  
\medskip\cr
\hline

22589   & $-$5& $<1$ & 5721.2 &  454 & 0.24 & 0.17$^a$& 0.0042 & 0.0025 & 
  0.22& 0.37 & 19 & 1.1& $-$28.0  &  0.6  & $-$0.03 & 1\cr

26886   &$-$8& 1 & 1263.2   &  3.7& 0.39 & 0.02& 0.025 & 0.002 &
  0.40& 0.32 & 23 & 2.0& +3.8   &  0.1  &  +0.02 & 2 \cr

27271   & $-$8& 1 & 1693.8   &  9.1& 0.22 & 0.02& 0.024 & 0.001 &
  0.31& 0.30 & 23 & 1.2& $-$18.1  &  0.1  & $-$0.01 & 2 \cr

40430    &  0& 1 & \multicolumn{1}{l@{\mbox{}}}{$>3700$} 
        &       & \multicolumn{1}{l@{\mbox{}}}{}      
        &       & \multicolumn{1}{l@{\mbox{}}}{}
        &       
   &      &  0.34 & 15 & 1.7& $-$23.9 & 1.0  &  $-0.03$ & 1 \cr

49841   &$-$8& 1 &  897.1   &  1.8& 0.16 & 0.01& 0.032 & 0.002 &
  0.33& 0.32 & 21 & 1.0& +10.9  &  0.1  & $-$0.09 & 2 \cr   

51959$^g$& 2& 1 & \multicolumn{1}{l@{\mbox{}}}{$>3700$} 
        &       & \multicolumn{1}{l@{\mbox{}}}{}      
        &       & \multicolumn{1}{l@{\mbox{}}}{}
        &       
   &      &  0.34 & 21 &  $-$ & +38.9 &  0.8  &  +0.07 & 1 \cr

53199   & $-$8& 2 & \multicolumn{1}{l@{\mbox{}}}{$7500$} & 
                                  & 0.21 & 0.22$^a$& 0.026 & 0.001 & 
  0.17& 0.36 & 11 & 1.1& +23.3  &  0.1  & $-$0.07 & 2\cr
         
58121   &  0& 1 & 1214.3   &  5.7& 0.14 & 0.02& 0.015 & 0.001 &
  0.24& 0.30 & 23 & 1.3& +10.2  &  0.1  & $-$0.04 & 2 \cr

58368   &  0& 2 &  672.7   &  1.3& 0.22 & 0.02& 0.021 & 0.001 &
0.39 & na   & 31 & na  & +37.8  &  0.1  & $-$0.11 & 13 \cr   

59852   & $-$9& 1&3463.9   & 53.8& 0.15 & 0.06& 0.0022& 0.0004& 
0.27 & 0.34 & 19 &  $-$ &  +0.1 & 0.1 & $-$0.10 & 1 \cr

77247$^{c,d}$&$-$5& 1 &   80.53  & 0.01& 0.09 & 0.01& 0.0050& 0.0001&
 0.50 & na  & 66  & 5.8  &$-$19.7 &  0.1   & na    & 0,13 \cr  

91208   &  0& 1 & 1754.0   & 13.3& 0.17 & 0.02& 0.022 & 0.002 &
  0.39& 0.32 & 24 & 0.7& +0.2   &  0.1  & $-$0.05 & 1 \cr 

95193   &  0& 1 & 1653.7   &  9.0& 0.13 & 0.02& 0.026 & 0.001 &
  0.26& 0.32 & 18 & 0.8& $-$7.3   &  0.1  & $-$0.06 & 1 \cr

98839$^b$&-7&$<1$& \multicolumn{1}{l@{\mbox{}}}{$>11000$} 
        &       & \multicolumn{1}{l@{\mbox{}}}{}      
        &       & \multicolumn{1}{l@{\mbox{}}}{}
        &       
   &      &       & 56 & na & \multicolumn{1}{c@{\mbox{}}}{$-$1:} &    &  na  & 5 \cr

101079  &  1& 1 & \multicolumn{1}{l@{\mbox{}}}{$>1500$}  
        &       & \multicolumn{1}{l@{\mbox{}}}{}      
        &       & \multicolumn{1}{l@{\mbox{}}}{}
        &       
   &      & 0.34 & 7  & 1.2& $-$2.2   &  0.2  & $-$0.17  & 2 \cr

104979  &  0& 1 & \multicolumn{1}{l@{\mbox{}}}{$>4700$} 
        &       & \multicolumn{1}{l@{\mbox{}}}{}      
        &       & \multicolumn{1}{l@{\mbox{}}}{}
        &       
   &      &  0.29 & 25 & 0.7& $-$30.8 &  0.4  & +0.07 & 2 \cr

131670$^{c,e}$&  1& 1 & 2929.7   & 12.2& 0.16 & 0.01& 0.040 & 0.002 &
  0.36& 0.30 & 55 & 0.6& $-$25.1  &  0.1  & $-$0.13 & 1,13 \cr 

134698  &  1& 1 & \multicolumn{1}{l@{\mbox{}}}{$>3600$}  
        &       & \multicolumn{1}{l@{\mbox{}}}{}      
        &       & \multicolumn{1}{l@{\mbox{}}}{}
        &       
   &      &  0.33 & 22 & 0.5& $-$29.5  &  1.6  & $-$0.20 & 1 \cr

139195  &  1& 1 & 5324     & 19  & 0.35 & 0.02& 0.026 & 0.002 &
 0.7  & na   &107 &  $-$  & +6.3   & 0.1  & $-$0.02 & 12 \cr

143899  &$-$8& 1 & 1461.6   &  6.9& 0.19 & 0.02& 0.017 & 0.001 &
  0.27& 0.35 & 26 & 0.3& $-$29.8  &  0.1  & $-$0.05 & 1 \cr    

165141  &  0& 1 & \multicolumn{1}{l@{\mbox{}}}{$> 2100$} 
        &       & \multicolumn{1}{l@{\mbox{}}}{}      
        &       & \multicolumn{1}{l@{\mbox{}}}{}
        &       
    &      &      & 10 & na &+10.2    & 1.8  & $-0.14$ & 4,23 \cr
180622  &  1& 1 & 4049.2 & 37.7 &   0.06 & 0.10$^a$ & 0.070 & 0.020       
   &  0.25  &  0.30 & 10  &  $-$ & +39.2  &  1.1  & $-$0.10  & 2 \cr 

196673$^{c,d}$&  2& 2 &  \multicolumn{1}{l@{\mbox{}}}{$ 6500$} &
                                 & 0.64 & 0.03& 0.013 &0.002 &
  0.47& 0.5   & 51 & 3.0 & $-$24.6   & 0.1  &  0.00 & 1,13 \cr

199394$^{c,d}$& $-$8& 1 & 4606.5 &351$^f$  & 0.06 & 0.06$^a$ & 
                                                0.023 & 0.003 &
  0.40&  na  & 52 & 1.1  & $-$5.6   & 0.2  & $-$0.15 & 0,13 \cr

200063  &  3& 1   & 1735.4   & 8.1 & 0.07 & 0.04$^a$& 0.058 & 0.004 &
  0.23& 0.29 & 10  & 2.3& $-$58.3 &  0.2 & $-$0.10  & 2 \cr

202109$^g$&$-$8& 1& 6489.0   & 31.0& 0.22 & 0.03& 0.023 & 0.003 &
  0.8 & na   &112 & 0.0& +16.7  & 0.1  &$-$0.06  & 11\cr

204075$^{c,d}$  & $-$5& 2 & 2378.2   & 55  & 0.28 & 0.07& 0.004 & 0.001 &
  0.52& na   & 32 & 4.9 &  +2.1   &  0.1  & $-$0.15 & 0,13 \cr  
 
205011$^{c,d}$  &  1& 1 & 2836.8     & 10  & 0.24 & 0.02& 0.034 & 0.003 &
  0.45& na   & 41 & 1.4 &  +11.5 &   0.1 & $-$0.13 & 0,13 \cr  

210946  &  1& 1 & 1529.5   &  4.1& 0.13 & 0.01& 0.041 & 0.001 &
  0.26& 0.31 & 30 & 1.3&$-$4.3   &  0.1  & $-$0.11 & 2 \cr

216219  & $-$1& 1 & 4098.0   &111.5& 0.10 & 0.04& 0.013 & 0.001 &
  0.37& 0.33 & 29 & 2.0& $-$7.2   &  0.1  & $-$0.07 & 2 \cr

223617$^{c,d,g}$&  2& 2 & 1293.7     &  3.9  & 0.06 & 0.02& 0.0064 & 0.0004 &
  0.34&  na  & 39 & 0.4  & +28.5  &  0.1  & $-$0.10 & 2,13 \cr     

288174  &  0& 1 & 1824.3   &  7.1& 0.19 & 0.01& 0.017 & 0.001 &
  0.15& 0.33 & 14 & 0.9& +34.7  &  0.1  & $-$0.12 & 1 \cr  
     
$-01^\circ3022$&  1& 1 & 3252.5 & 31.4 & 0.28 & 0.02& 0.016 & 0.001 & 
  0.25 &  0.36 & 26 & 1.0& $-$35.4  &  0.1  & $-$0.15 & 1 \cr
      
$-10^\circ4311$& $-$0& 1 & \multicolumn{1}{l@{\mbox{}}}{$>3400$}  
        &       & \multicolumn{1}{l@{\mbox{}}}{}      
        &       & \multicolumn{1}{l@{\mbox{}}}{}
        &       
   &      &  0.45 & 33 & 1.5& +52.7 &  2.9  & $-$0.14 & 1 \cr

$-14^\circ2678$&  0&$<1$ & 3470.5 & 107 & 0.22 & 0.04& 0.023 & 0.002 &
  0.39 &  0.38 & 15 & 2.9& +4.9   &  0.1  & +0.06 & 1 \cr 
\hline
\end{tabular}

Remarks:\\
a: data compatible with circular orbit at 5\% confidence level
(Lucy-Sweeney test);\\
b: not listed in L\"u et al. (1983), but present in L\"u (1991);\\  
c: Combined CORAVEL/DAO orbit;\\ 
d: A DAO-CORAVEL offset of $-$0.46~\kms\ has been applied to the DAO 
measurements;\\
e: A DAO-CORAVEL offset of $-$0.73~\kms\ has been applied to the DAO 
measurements;\\
f: A somewhat more accurate period ($4382\pm91$~d) is obtained by
forcing $e = 0$;\\
g: Acceleration solution listed in the Hipparcos Double and Multiple
Systems Annex (ESA 1997) 
}
\end{table*}

\clearpage
\newpage

\addtocounter{table}{-1}
\renewcommand{\thetable}{\arabic{table}b}
\begin{table*}
\caption[]{
Suspected binary mild barium stars. The numbers in column `Ref.'
refer to the papers listed in Table~5
}
\small{
\begin{tabular}{lllrcrccccll}
\hline\cr
HD & Sp. & Ba & $N$ & $\Delta t$ & \multicolumn{1}{c@{\mbox{}}}{$V_r$} 
& $\sigma(V_r)$ & \multicolumn{1}{c@{\mbox{}}}{$\overline{\epsilon}_1$} & \multicolumn{1}{c@{\mbox{}}}{$S_b$}
 & $\Delta(38-41)$ & Ref.
& Rem.\cr
     &     &    &     & (d)          & (km/s)          & (km/s)               &
(km/s)     & (km/s)
\medskip\\
\hline
18182    &  0& 0 & 22 & 3451 & 25.76   & 0.45&  0.32& $-$ & $-$0.03 & 0 & \cr
183915   &  0& 2 & 9  & 4073 &$-$50.08 & 0.50&  0.29& 0.3 & $-$0.18 & 0,13 & \\
218356   &  2& 2 & 13 & 6201 &$-$27.86 & 1.17&  0.29& 3.6 & $-$0.13 & 0,19 & 
56 Peg (K0IIp + WD)\\
\hline
\end{tabular}
}
\end{table*}

\addtocounter{table}{-1}
\renewcommand{\thetable}{\arabic{table}c}
\begin{table*}
\caption[]{
Mild barium stars with no evidence of binary motion. The numbers in column `Ref.'
refer to the papers listed in Table~5
}
\small{
\begin{tabular}{lllrcrccccll}
\hline\cr
HD & Sp. & Ba & $N$ & $\Delta t$ & \multicolumn{1}{c@{\mbox{}}}{$V_r$} &
$\sigma(V_r)$ & \multicolumn{1}{c@{\mbox{}}}{$\overline{\epsilon}_1$} & \multicolumn{1}{c@{\mbox{}}}{$S_b$}
&  $\Delta(38-41)$ & Ref.
& Rem.\cr
     &     &    &     & (d)          & (km/s)          & (km/s)               &
(km/s)     &  (km/s)
\medskip\cr
\hline
50843    &  1& 1 & 20 & 4437 &  12.21 &  0.33&  0.33& $-$&  $-$0.03 & 0 & \cr
95345    &  2& 1 & 26 & 3036 &  5.48  &  0.21&  0.28& 0.7&    +0.07 & 0 & \cr
119185   &  0& 1 & 17 & 2869 &  $-$74.51&0.24&  0.36& 0.8&  $-$0.05 & 0 & \cr
130255   &  0& 1 & 26 & 2956 &  40.19 &  0.33&  0.32& $-$&    +0.01 & 0,9 & subgiant CH\cr
\hline
\end{tabular}
}
\end{table*}

\addtocounter{table}{-1}
\renewcommand{\thetable}{\arabic{table}d}
\begin{table*}
\caption[]{
Supergiants misclassified as mild barium stars. The numbers in column `Ref.'
refer to the papers listed in Table~5
}
\small{
\begin{tabular}{lllrcrccccll}
\hline\cr
HD  & Sp. & Ba & $N$ & $\Delta t$ & \multicolumn{1}{c@{\mbox{}}}{$V_r$} 
& $\sigma(V_r)$ & \multicolumn{1}{c@{\mbox{}}}{$\overline{\epsilon}_1$} 
& \multicolumn{1}{c@{\mbox{}}}{$S_b$}
& $\Delta(38-41)$ & Ref.
& Rem.\cr
     &     &    &     & (d)          & (km/s)          & (km/s)               &
(km/s)   & (km/s)   
\medskip\cr
\hline
65699 & 0& 0 & 21 & 5163 & 11.17  & 0.27&  0.31& 6.0& $-$0.01 & 0,16 & \cr
206778& 2& 1 & 94 & 6204 & 2.82   & 0.56&  0.27& 5.8&   +0.06 & 0,16 & 
$\epsilon$ Peg (K2II/Ib var)\\
\hline
\end{tabular}
}
\end{table*}

\clearpage
\newpage

\renewcommand{\thetable}{\arabic{table}a}

\begin{table*}
\caption[]{\label{Tab:orbiteBas}
Same as Table~\protect\ref{Tab:orbiteBam} for strong (Ba3, Ba4 and Ba5) barium 
stars. The numbers in column `Ref.'
refer to the papers listed in Table~5
}
\tabcolsep 3pt
{\small
\begin{tabular}{lrll@{$\pm$}ll@{$\pm$}ll@{$\pm$}llllcr@{$\pm$}lcl}
\hline\cr
 & & & \multicolumn{2}{c@{\mbox{}}}{}  &
\multicolumn{2}{c@{\mbox{}}}{}      & 
\multicolumn{2}{c@{\mbox{}}}{}   &
& &  & & 
\multicolumn{2}{c@{\mbox{}}}{} &
$\Delta$\cr
HD/DM/ & Sp. & Ba & 
\multicolumn{2}{c@{\mbox{}}}{$P$}      &
\multicolumn{2}{c@{\mbox{}}}{$e$}      & 
\multicolumn{2}{c@{\mbox{}}}{$f(M)$}   &
$O-C$                                  & 
\multicolumn{1}{c@{\mbox{}}}{$\overline{\epsilon}_1$}& 
$N$                                    & 
\multicolumn{1}{c@{\mbox{}}}{$S_b$}    & 
\multicolumn{2}{c@{\mbox{}}}{$\gamma$} &
$(38-41)$ & Ref. \cr
others & & &
\multicolumn{2}{c@{\mbox{}}}{(d)}      &
\multicolumn{2}{c@{\mbox{}}}{}      & 
\multicolumn{2}{c@{\mbox{}}}{(\Msun)}   &
(km/s) & (km/s) & & (km/s) & 
\multicolumn{2}{c@{\mbox{}}}{(km/s)} &  &  
\medskip\cr
\hline

5424    &  1& 4 & 1881.5   & 18.6& 0.23 & 0.04& 0.005 & 0.0004&
  0.18& 0.30 & 13 & $-$  &  $-$0.3 & 0.1 & $-$0.11 & 1 \cr 

16458$^g$&  1& 5 & 2018     & 12  & 0.10 & 0.02& 0.041 & 0.003 &
  0.38&  na  & 36 & na   &+20.3  & 0.1 & $-$0.13 & 13 \cr 

20394   &  0& 4 & 2226     & 22  & 0.20 & 0.03& 0.0020& 0.0002 &
 0.34 & -   & 87 & $-$ & +24.2 & 0.1 & $-$0.09 & 6 \cr
                       
24035   &  4& 4 &  377.8   &  0.3& 0.02 & 0.01$^a$
                                              & 0.047 & 0.003 &
  0.19& 0.29 & 15 & $-$ & $-$12.5 & 0.1 & $-$0.26 & 1\cr

31487   &  1& 5 & 1066.4   &  2.6& 0.05 & 0.01& 0.038 & 0.002 &
  0.33&  na  & 35 & na  &  $-$4.2 & 0.7 & na    & 13\cr

36598   &  2& 4 & 2652.8   & 22.7& 0.08 & 0.02& 0.037 & 0.002 &
  0.21& 0.27 & 11 & $-$  & +44.1 & 0.1 & $-$0.18 & 1 \cr 

42537   &  4& 5 & 3216.2   & 54.7& 0.16 & 0.05
                                              & 0.027 & 0.005 &
  0.43& 0.30 & 12 & 1.9&  $-$2.5 & 0.2 & $-$0.31 & 1\cr 

43389   &  2& 5 & 1689.0   &  8.7& 0.08 & 0.02& 0.043 & 0.002 &
  0.35& 0.32 & 24 & 0.6& +53.1 & 0.1 & $-$0.16 & 2\cr   

44896   &  3& 5 &  628.9   &  0.9& 0.02 & 0.01$^a$
                                              & 0.048 & 0.001 &
  0.21& 0.26 & 19 & 1.9& +52.2 & 0.1 & $-$0.17 & 2\cr

46407$^{b,e}$&0&3 &  457.4   &  0.1& 0.013 & 0.008$^a$ 
                                              & 0.035 & 0.001 &
  0.40& 0.29 & 68 & 1.8&  $-$3.4 & 0.1 & $-$0.21 & 2,13\cr

49641   &  1& 3 & 1768     & 23  & \multicolumn{2}{c@{\mbox{}}}{0.0} & 
                                                0.0031& 0.0004 &
 0.42 & na & 35 & na  & +4.4  &  0.1 & $-$0.14 & 13 \cr

50082   &  0& 4 & 2896.0   & 21.3& 0.19 & 0.02& 0.027 & 0.002 &
  0.35& 0.31 & 29 & 1.1 & $-$17.4 & 0.1 & $-$0.12 & 2 \cr

60197   &  3& 5 & 3243.8   & 66.3& 0.34 & 0.05& 0.0028& 0.0006&
  0.31& 0.27 & 14 & 3.0 & +54.3 & 0.1 & $-$0.05 & 1\cr

84678   &  2& 4 & 1629.9   & 10.4& 0.06 & 0.02$^a$
                                              & 0.062 & 0.003 &
  0.30& 0.29 & 12 & 0.8& +27.9 & 0.1 & $-$0.31 & 1\cr  

88562   &  2& 4 & 1445.0   &  8.5& 0.20 & 0.02& 0.048 & 0.003 &
  0.44& 0.32 & 23 & 0.7& +11.8 & 0.1 & $-$0.04 & 1\cr  

92626   &  0& 5 &  918.2   &  1.2& 0.00 & 0.01$^a$
                                              & 0.042 & 0.002 &
  0.32& 0.27 & 35 & 0.5& +16.3 & 0.1 & $-$0.29 & 2\cr  

100503  &  3& 5 &  554.4   &  1.9& 0.06 & 0.05$^a$
                                              & 0.011 & 0.001 &
  0.55& 0.28 & 16 & 1.7&  $-$8.9 & 0.1 & $-$0.22 & 1\cr  

101013$^e$&  0& 5 & 1711   &  4  & 0.20 & 0.01& 0.037 & 0.001 &
  0.58& na   &118 & 1.4& $-$14.5 & 0.1 & $-$0.17 & 13,20 \cr 

107541  &  0& 4 & 3569.9   & 46.1& 0.10 & 0.03
                                              & 0.029 & 0.002 &
  0.28& 0.31 & 16 & 0.1 & +88.1 & 0.1 & $-$0.26 & 2 \cr

120620  &  0& 4 &  217.2   &  0.1& 0.01 & 0.01$^a$
                                              & 0.062 & 0.001 &
  0.42& 0.36 & 28 &  $-$  & +33.2 & 0.1 & $-$0.18 & 1 \cr                    

121447  &  7& 5 &  185.7   &  0.1& 0.015 & 0.013$^a$
                                              & 0.025 & 0.001 &
  0.47& 0.32 & 26 & 2.8&$-$11.9 & 0.1 &$-$0.26 & 7 \cr 

123949  &  6& 4 & \multicolumn{2}{l@{\mbox{}}}{9200}
                & 0.97 & 0.06 & 0.105 & 0.64 &
  0.30 & 0.33 & 25 & 0.6& $-$10.8 & 0.3 & $-$0.19 & 1\\  
  
154430  &  2& 4 & 1668.1   & 17.4& 0.11 & 0.03$^a$
                                              & 0.034 & 0.003 &
  0.48& 0.29 & 15 & 1.3& $-$38.1 & 0.1 & $-$0.04 & 1\cr  

178717  &  4& 5 & 2866     & 21  & 0.43 & 0.03& 0.006 & 0.001 &
  0.47&  na  & 46 & 2.0  & $-$16.4 & 0.1  & $-$0.20 & 13 \cr 

196445  &  2& 4 & 3221.3   & 43.0& 0.24 & 0.02& 0.031 & 0.002 &
  0.23& 0.29 & 12 & 0.6 & $-$25.5 & 0.1 & $-$0.22 & 1 \cr

199939  &  0& 4 &  584.9   &  0.7& 0.28 & 0.01& 0.025 & 0.001 &
  0.47&  na  & 52 & 1.6 &  $-$41.7 & 0.1 & $-$0.25 & 13\cr  

201657  &  1& 4 & 1710.4   & 15.0& 0.17 & 0.07& 0.004& 0.001&
  0.29&  0.31& 15 & $-$&  $-$27.7 & 0.2  &  $-$0.24 & 2\cr 

201824  &  0& 4 & 2837     & 13  & 0.34 & 0.02& 0.040 & 0.003 &
 0.45 & -   & 86 & 0.8& $-$31.1 & 0.1  & $-$0.18 & 6 \cr  

211594  &  0& 4 & 1018.9   &  2.7& 0.06 & 0.01& 0.0140& 0.0005 &
  0.33& 0.30 & 49 & 1.0&  $-$9.9 & 0.1 & $-$0.39 & 2\cr      
  
211954  &  2& 5 & \multicolumn{1}{l@{\mbox{}}}{5000}  
        &       & 0.39 & 0.08 & 0.017 & 0.005 &
  0.35 & 0.32 & 14  & $-$ &  $-$6.1 & 0.1 & $-$0.22 & 1\cr

$+38^\circ118$(a+b)$^c$&2&5&299.4&  0.2& 0.14 & 0.01& 0.0141& 0.0004&
  0.30& 0.31 & 30 & 1.4& $-$18.7 & 0.1 & $-$0.19 & 1 \cr  

$+38^\circ118$(ab+c)$^c$&2&5&3876.7&112.2& 0.21 & 0.06& 0.0017 & 0.0004&
  0.29& 0.31 & 30 & 1.4& $-$18.3 & 0.1 & $-$0.19 & 1 \cr    

$-42^\circ2048$&  2& 4 & 3260.0   & 28.3& 0.08 & 0.02& 0.065 & 0.004 &
  0.24& 0.29 & 12 & 1.4 & +40.5 & 0.1 & $-$0.16 & 1 \cr

$-64^\circ4333^d$&0&4 &  386.0   &  0.5& 0.03 & 0.01$^a$
                                              & 0.068 & 0.003 &
  0.31& 0.33 & 16 & $-$  &  +8.3 & 0.2 & $-$0.29 & 1\cr

L\"u 163 & $-$5& 5 &  965.1   & 16.0& 0.03 & 0.07$^a$
                                              & 0.0029 & 0.0006&
  0.57& 0.39 & 14 &  $-$ &  +2.8 & 0.2 & $-$0.39 & 1\cr

NGC 2420 X & - & 5 & 1402 & 10 & \multicolumn{2}{c@{\mbox{}}}{0.0}
                                              & 0.050 & 0.005 &
  0.50 & na  & 16 & $-$  & +78.2 & 0.2 & na     & 13,21\cr
\hline
\end{tabular}

Remarks:\\
a: data compatible with circular orbit at 5\% confidence level
(Lucy-Sweeney test);\\
b: Combined CORAVEL/DAO orbit; A DAO-CORAVEL offset of $-$0.19~\kms\ has been applied to the DAO 
measurements;\\
c: triple system;\\ 
d: CpD;\\
e: Orbital solution listed in the Hipparcos Double and Multiple
Systems Annex (ESA 1997);\\ 
g: Acceleration solution listed in the Hipparcos Double and Multiple
Systems Annex (ESA 1997) 
}
\end{table*}

\clearpage
\newpage

\addtocounter{table}{-1}
\renewcommand{\thetable}{\arabic{table}b}
\begin{table*}
\caption[]{
Strong barium stars with no evidence for binary motion. The numbers in column `Ref.'
refer to the papers listed in Table~5
}
\small{
\begin{tabular}{cccccrcccccl}
\hline\cr
HD & Sp. & Ba & $N$ & $\Delta t$ & 
\multicolumn{1}{c@{\mbox{}}}{$V_r$} & 
$\sigma(V_r)$ & 
\multicolumn{1}{c@{\mbox{}}}{$\overline{\epsilon}_1$} & \multicolumn{1}{c@{\mbox{}}}{$S_b$} &
$\Delta(38-41)$ & Ref.
& Rem.\cr
     &     &    &     & (d)          & (km/s)          & (km/s)               &
(km/s)   & (km/s)
\medskip\cr
\hline
19014    &  4& 5 & 18 & 3272 & 13.3 & 0.47&  0.28& 2.2& $-$0.01 & 0  &jitter only?\cr
65854    &  1& 3 & 30 & 3369 & 0.5 & 0.42&   na & na &   na    & 13  & \cr
\hline
\end{tabular}

}
\end{table*}

\tabcolsep 2pt
\renewcommand{\thetable}{\arabic{table}a}
\begin{table*}
\caption[]{\label{Tab:orbiteS}
Orbital elements of S stars. 
Column 2, labeled GCGSS, lists the 
star number in the {\it General Catalogue of Galactic S Stars}
(Stephenson 1984). The numbers in column `Ref.' and `Ref. Tc'
refer to the papers listed in Table~5
}
\small{
\begin{tabular}{l@{\hspace{0pt}}rll@{$\pm$}l@{\hspace{6pt}}l@{$\pm$}ll@{$\pm$}llllcr@{$\pm$}@{\extracolsep{2pt}}lllll}
\hline\cr
HD/DM & {\small GCGSS} & Sp. & 
\multicolumn{2}{c@{\mbox{}}}{$P$}      &
\multicolumn{2}{c@{\mbox{}}}{$e$}      & 
\multicolumn{2}{c@{\mbox{}}}{$f(M)$}   &
$O-C$                                  &
\multicolumn{1}{c@{\mbox{}}}{$\overline{\epsilon}_1$}& 
$N$                                    & 
\multicolumn{1}{c@{\mbox{}}}{$S_b$}    & 
\multicolumn{2}{c@{\mbox{}}}{$\gamma$} & 
Ref. & Tc & Ref. & Rem.\cr
     &                &                &
\multicolumn{2}{c@{\mbox{}}}{(d)}      &
\multicolumn{2}{c@{\mbox{}}}{}         & 
\multicolumn{2}{c@{\mbox{}}}{(\Msun)}  &
(km/s) & (km/s) &  & (km/s)            & 
\multicolumn{2}{c@{\mbox{}}}{(km/s)}   &
orb. &    & Tc &\cr
\hline

7351& 26 & S3/2 & 4593 & 110 & 0.17 & 0.03 & 0.073 & 0.007 & 
0.68 &  0.31 & 50 & 3.3&+1.5 & 0.1 & 3,4 & n &15 & HR 363\cr

22649$^f$& 79 & S4/2&  596.2&  0.2 & 0.09 & 0.02  & 0.037 & 0.003& 
0.8 & na  & 53& na  & $-$22.3   & 0.1  & 18  & n &15 & HR 1105\cr

30959$^d$&  114 & S3/1 & \multicolumn{1}{l@{\mbox{}}}{$>1900$} 
                                                  &       &
\multicolumn{1}{l@{\mbox{}}}{} &       &
\multicolumn{1}{l@{\mbox{}}}{}       &       
&   & 0.29 & 12 & 4.3&  $-$8.8 & 0.7 & 1,14 &y  & 15 & $o^1$ Ori\cr

35155&  133 & S4,1& 640.5  &  2.8   & 0.07 & 0.03  & 0.032 & 0.003& 
0.81$^c$& 0.33& 19& 3.5& +79.7 & 0.2 & 0,10 &n & 15 & \\

246818 &  156 & S   & 2548.5&73.2 & 0.18 & 0.11$^a$ 
                                                   & 0.0035& 0.0015&
0.59& 0.37& 17& 2.8& $-$45.5 & 0.2 & 1 & n & 8 & $+05^\circ1000$\cr

288833& 233 & S3/2 &  \multicolumn{1}{l@{\mbox{}}}{$>3900$} 
                                                  &       &
\multicolumn{1}{l@{\mbox{}}}{} &       &
\multicolumn{1}{l@{\mbox{}}}{}       &       
&   & 0.38 & 18 & 3.5&  +81.1 & 1.0 & 1  & n & 8 & $+02^\circ1307$\cr

49368 &  260 & S3/2& 2995.9 & 67.1 & 0.36 & 0.05 & 0.022 & 0.003 &
0.58  &   0.34& 23& 4.3& +49.8 & 0.1 & 1 & n & 15 & V613 Mon\cr

63733   &  411 & S4/3& 1160.7&  8.9 & 0.23 & 0.03  & 0.025 & 0.003& 
0.38& 0.32& 14& 3.5& +1.9  & 0.1 & 1 &y?& 15 & \cr

95875   & 720  & S3,3&  197.2&  0.4 & 0.02 & 0.04$^a$& 0.059 & 0.009 &
0.70& 0.28 & 10 & 4.0 & +40.6 & 1.1 & 1 & n& 0 & Hen 108 \cr

121447  &  -   & S0  &  185.7&  0.1 & 0.015 & 0.013$^a$
                                                   & 0.025 & 0.001 &
0.47& 0.32& 26& 2.8& $-$11.9 & 0.1 & 7 &n &17 &\cr 
   
170970  & 1053 & S3/1& 4392& 202& 0.08 & 0.04$^a$
                                                   & 0.021 & 0.002& 
0.33& 0.30& 39& 3.9& $-$35.8 & 0.1 & 1 &y?&15 & \cr

184185& 1140 & S3*4 &  \multicolumn{1}{l@{\mbox{}}}{$>3400$} 
                                                  &       &
\multicolumn{1}{l@{\mbox{}}}{} &       &
\multicolumn{1}{l@{\mbox{}}}{} &       
&   & 0.43 & 22 & 4.3&  +1.5 & 1.6 & 1 & - & - & $-21^\circ5435$\\
   
191226  & 1192 & M1S & 1210.4&  4.3 & 0.19 & 0.02  & 0.013 & 0.001& 
0.38& 0.30& 36& 4.2& $-$25.0 & 0.1 & 3 & n &15 & \cr
          
191589  & 1194 & S   &  377.3&  0.1 & 0.25 & 0.003 & 0.394 & 0.005& 
0.29& 0.30& 41& 3.1&  $-$9.7 & 0.1 & 1 &n & 15 &\cr

218634$^e$& 1322 & M4S &  \multicolumn{1}{l@{\mbox{}}}{$>3700$} 
                                                  &       &
\multicolumn{1}{l@{\mbox{}}}{} &       &
\multicolumn{1}{l@{\mbox{}}}{}       &       
&   & 0.36 & 28 & 5.6& +20.1 & 2.3 & 0,22 & n & 8 & 57 Peg\cr

332077  & 1201 & S3,1&  669.1&  1.0 & 0.077 & 0.007 & 1.25 & 0.02& 
0.66& 0.46& 39&10.2&$-$5.2 & 0.1  & 0,10 &n & 8 & \\

343486  & 1092 & S6,3& 3165.7& 37.6 & 0.24 & 0.03  & 0.039 & 0.005& 
0.82& 0.43& 37& 3.8&  +4.9 & 0.1 & 1 & - & -  &\cr

$+21^\circ255^b$& 45 & S3/1& 4137 & 317 & 0.21 & 0.04 & 0.032 & 0.004&
0.51 & 0.33& 36& 3.2& $-$38.5 & 0.3 & 1 & n & 8 & \cr

$+24^\circ620$ &  87  & S4,2&  773.4&  5.5 & 0.06 & 0.03$^a$ & 0.042 & 0.005& 
0.82& 0.41& 19& 4.0& $-$21.0 & 0.2 & 0,10 &n & 8 & \\

$+22^\circ700$ &  96  & S6,1&  849.5 & 8.8 & 0.08 & 0.06$^a$
                                                   & 0.043 & 0.008& 
1.14& 0.48& 20& 4.9& +40.5 &  0.3 & 0,10 &n & 8 & \\

$+79^\circ156$ &  106 & S4,2& \multicolumn{1}{l@{\mbox{}}}{$>3900$} 
                                                  &       &
\multicolumn{1}{l@{\mbox{}}}{}&       &
\multicolumn{1}{l@{\mbox{}}}{}&       
   &   & 0.39 & 19 & 4.3& $-$33.0 & 2.1 & 1 & n & 8 & \cr

$+23^\circ3093$&  981 & S5,4& 1008.1  & 4.8  & 0.39 & 0.03  & 0.045 & 0.005& 
0.81& 0.41& 30& 3.8& $-$44.1 & 0.2 & 0,10 &n & 8 & \\

$+23^\circ3992$& 1209 & S3,3& 3095.6& 41.7 & 0.10 & 0.03$^a$
                                                   & 0.034 & 0.004& 
0.71& 0.37& 43& 4.3& $-$26.7 & 0.1 & 1 & n & 8 & \cr

$+31^\circ4391$& 1267 &S2/4 & \multicolumn{1}{l@{\mbox{}}}{$>3600$}
                                                  &       &
\multicolumn{1}{l@{\mbox{}}}{} &       &
\multicolumn{1}{l@{\mbox{}}}{}       &       
&   & 0.37& 28& 3.3& +25.2 & 1.6 & 1 & - & -&\\

$+28^\circ4592$& 1334 & S2/3:&1252.9&  3.5 & 0.09 & 0.02  & 0.016 & 0.001& 
0.32& 0.34& 34& 3.2& $-$37.5 & 0.1 & 1 & n & 8 & \cr
\hline
\end{tabular}

Remarks:\\
a: Data compatible with circular orbit at 5\% confidence level
(Lucy-Sweeney test);\\
b: Visual binary; the S star is BD+21$^\circ$255
= PPM 91178 = SAO 75009 = HIC 8876, whereas its visual K-type
companion (BD+21$^\circ$255p = PPM 91177 = SAO 75008) is also a
spectroscopic binary whose orbit is given in Jorissen \& Mayor (1992);\\ 
c: Two outlying measurements (at phase 0.38 and 0.93 deviating by
1.40 and $-1.98$~\kms, respectively) were kept in the present
orbital solution. No obvious instrumental origin could be found to
account for these outlying measurements, which may have a real - but
as yet unidentified - physical cause in this strongly interacting
system (Ake et al. 1991);\\
d: The WD companion of $o^1$ Ori has been detected by IUE (Ake \& Johnson 1988);\\ 
e: 57 Peg has a composite spectrum S+A6V (Van Eck et al. 1997);\\
f: Orbital solution listed in the Hipparcos Double and Multiple
Systems Annex (ESA 1997) 
}
\end{table*}

\clearpage
\newpage

\tabcolsep 6pt
\addtocounter{table}{-1}
\renewcommand{\thetable}{\arabic{table}b}
\begin{table*}
\caption[]{
Non-binary (misclassified?) S stars
}
\small{
\begin{tabular}{lrlccrcccllll}
\hline\cr
HD/DM          &GCGSS& Sp. &$N$&$\Delta t$
& \multicolumn{1}{c@{\mbox{}}}{$V_r$}  
& $\sigma(V_r)$ 
& \multicolumn{1}{c@{\mbox{}}}{$\overline{\epsilon}_1$} 
& \multicolumn{1}{c@{\mbox{}}}{$S_b$} 
& Tc & Ref. & Rem.\cr 
               &     &     &   & (d)      &    (km/s)     & (km/s)                
&(km/s) & (km/s)          &    & Tc   &      \cr
\hline\cr   

262427          &247  & S?  & 10 & 3622 & +34.61 & 0.37 &   0.31 &  2.3& -  & 
- & S class from Perraud (1959)  \cr
$+22^\circ4385$ &1271 & S2 & 26 & 3607 &  $-$2.63& 0.36 &   0.36 &  2.6& -  & 
- & S class from Vyssotsky \& Balz (1958) \cr
\hline
\end{tabular}
}
\end{table*}

\addtocounter{table}{-1}
\renewcommand{\thetable}{\arabic{table}c}
\begin{table*}
\caption[]{
S stars with radial-velocity jitter. The numbers in column `Ref. Tc'
refer to the papers listed in Table~5 
}
\small{
\begin{tabular}{lrlccrcccllll}
\hline\cr
HD/DM          &GCGSS& Sp. &$N$&$\Delta t$& 
\multicolumn{1}{c@{\mbox{}}}{$V_r$} & 
$\sigma(V_r)$ & 
\multicolumn{1}{c@{\mbox{}}}{$\overline{\epsilon}_1$} & 
\multicolumn{1}{c@{\mbox{}}}{$S_b$} & 
Tc & Ref. & Rem.\cr 
               &     &     &   & (d)      &    (km/s)     & (km/s)                
&(km/s) & (km/s)          &    & Tc   & \cr    
\hline\cr   
\BD{-10}{1334} & 176 & Sr   & 16 & 3613 &  +22.72 &1.39 &   0.48 & 7.1 & -  & - \cr     
\BD{+15}{1200} & 219 & S4/2 & 19 & 3933 &  +46.77 &1.47 &   0.45 & 5.2 & -  & - \cr       
61913          & 382 & M3S  & 12 & 1860 & $-$15.68&0.65 &   0.30 & 4.9 & dbfl & 17 & NZ Gem = HR 2967 \cr
\BD{-04}{2121} & 416 & S5/2 & 18 & 3585 &  +31.83 &1.04 &   0.38 & 5.6 &yes & 8 \cr   
\BD{-21}{2601} & 554 & S3*3 & 14 & 3308 &  +44.72 &1.07 &   0.36 & 3.5 & no  & 0 & \cr       
\BD{+20}{4267} &1158 & Swk  & 19 & 3009 &  +26.69 &1.59 &   0.41 & 6.9 & -  & - \cr    
189581         & 1178& S4*2 & 19 & 3399 & $-$17.04&0.78 &   0.34 & 4.5 & no & 15 & \cr
\BD{+04}{4354} &1193 & S4*3 & 13 & 3006 & $-$8.79 &0.63 &   0.39 & 5.3 &yes & 8 \cr   
192446         &1198 & S6/1 & 17 & 3245 & $-$22.48&0.79 &   0.42 & 6.0 &yes & 8 & \cr
216672         &1315 & S4/1 & 30 & 3251 &  +12.47 &0.76 &   0.31 & 4.9 &yes& 15 & HR Peg = HR 8714\cr
\hline
\end{tabular}
}
\end{table*}

\addtocounter{table}{-1}
\renewcommand{\thetable}{\arabic{table}d}
\begin{table*}
\caption[]{
Mira S stars. The numbers in column `Ref. Tc'
refer to the papers listed in Table~5
}
\small{
\begin{tabular}{lrlccrrccclll}
\hline\cr
HD           &GCGSS& Var & Sp. &$N$&$\Delta t$& 
\multicolumn{1}{c@{\mbox{}}}{$V_r$} & 
$\sigma(V_r)$ & 
\multicolumn{1}{c@{\mbox{}}}{$\overline{\epsilon}_1$} & 
\multicolumn{1}{c@{\mbox{}}}{$S_b$} & 
Tc & Ref. & Rem.\cr 
               &     &&     &   & (d)      &    (km/s)     & (km/s)                
&(km/s) & (km/s)         &    & Tc   &      \cr
\hline\cr   
1967       &  14& R And &S5-7/4-5e&9& 1924 & $-$4.85  & 4.92 &   0.54 & 9.6 &yes &17 &  \cr
4350       &  12& U Cas &S5/3e & 5  &  729 & $-$51.62 & 7.97 &   0.78 & 8.5 &yes &17 &  \cr
14028      &  49& W And & S7/1e& 3  & 1856 & $-$38.34 & 0.70 &   0.42 & 4.2 &yes &17 &  \cr
29147      & 103& T Cam & S6/5e& 3  &  641 & $-$11.09 & 5.28 &   0.55 &11.5 &yes &17 &  \cr
53791      & 307& R Gem & S5/5 & 7  & 2126 & $-$45.55 & 4.05 &   0.58 &11.7 &yes &17 &  \cr
70276      & 494& V Cnc & S3/6e& 4  & 2126 &  $-$4.71 &10.17 &   0.48 &10.3 &dbfl&17 &  \cr
110813     & 803& S UMa &S3/6e & 13 & 2667 &  +1.46   & 5.15 &   0.45 &10.1 &yes &17 &  binary?\cr  
117287     &  - & R Hya & M6e-M9eS&16&2924 &  $-$9.69 & 2.06 &   0.70 & 9.6 &prob&17 &  \cr
185456     &1150& R Cyg & S6/6e& 4  &  767 & $-$29.45 & 2.42 &   0.58 &13.2 &yes &17 &  \cr
187796     &1165&$\chi$ Cyg&S7/1.5e&12&2976&   +3.24  & 3.19 &   0.61 & 8.3 &yes &17 & \cr
190629     &1188& AA Cyg& S6/3 & 31 & 2931 & +11.41   & 1.33 &   0.42 & 8.7 &yes &17 &  \cr
195763     &1226& Z Del & S4/2e& 3  &  348 & +37.43   & 3.52 &   0.93 & 9.7 &yes &17 &  \cr
211610     &1292& X Aqr & S6,3e& 7  & 1746 & +11.67   & 7.20 &   0.75 & 8.3 &prob&17 &  \cr
\hline
\end{tabular}
}
\end{table*}

\clearpage
\newpage

\addtocounter{table}{-1}
\renewcommand{\thetable}{\arabic{table}e}
\begin{table*}
\caption[]{
SC Stars. The numbers in column `Ref.'
refer to the papers listed in Table~5
}
\small{
\begin{tabular}{lrlccrrccclll}
\hline\cr
HD/DM   &GCGSS& Var & Sp. &$N$&$\Delta t$& 
\multicolumn{1}{c@{\mbox{}}}{$V_r$} & 
$\sigma(V_r)$ & 
\multicolumn{1}{c@{\mbox{}}}{$\overline{\epsilon}_1$} & 
\multicolumn{1}{c@{\mbox{}}}{$S_b$} & 
Tc & Ref. & Rem.\cr 
               & &     &     &   & (d)      &    (km/s)     & (km/s)                
&(km/s) & (km/s)          &    & Tc   &      \cr
\hline\cr   
286340         & 117 & GP Ori &SC7/8 & 15 & 3614 &  +82.78 &2.09 &   0.45 & 7.0 & - & - &  \cr
44544          & 212 & FU Mon & S7/7 & 17 & 3720 & $-$25.86&1.71 &   0.48 & 7.2 & - & -&  \cr
\BD{-04}{1617} & 244 & V372 Mon&SC7/7& 13 & 3341 &  +16.66 &2.93 &   0.46 & 7.3 & -  & - &  \cr   
\BD{-08}{1900} & 344 &        & S4/6 & 17 & 3619 &  +72.47 &1.54 &   0.40 & 5.4 & -  & - & binary?\cr 
54300          &     & R CMi  & CS   & 16 & 3740 &  +45.76 &6.06 &   0.52 &12.2 &yes &17 &  pseudo-orbit\cr
198164         &     & CY Cyg & CS   & 27 & 3613 &   +4.52 &1.02 &   0.34 & 6.9 &yes&17 &  \cr    
209890         &     & RZ Peg & C9   & 21 & 2932 & $-$32.32&6.51 &   0.56 &14.9 &yes &17 &  pseudo-orbit\cr
\hline\cr                                                             
\end{tabular}

}
\end{table*}

\renewcommand{\thetable}{\arabic{table}}
\begin{table*}
\caption[]{
Tc-poor carbon stars. The column labelled GCCCS refers to the entry
number in the  {\it General Catalogue of Cool Carbon Stars}
(Stephenson 1973)  
}
\small{
\begin{tabular}{lrllrrrcccllll}
\hline\cr
HD           & GCCCS & Var & Sp. &$N$&$\Delta t$& 
\multicolumn{1}{c@{\mbox{}}}{$V_r$} & 
$\sigma(V_r)$ & 
\multicolumn{1}{c@{\mbox{}}}{$\overline{\epsilon}_1$} & 
\multicolumn{1}{c@{\mbox{}}}{$S_b$} & 
Rem.\cr 
             &     &       &     &   & (d)      &    (km/s)     & (km/s)                
&(km/s) & (km/s)          &    &    &      \cr
\hline\cr   
46687  &  537 & UU Aur & C5,3 & 6 & 857 &  13.5 & 1.27 & 0.28 & 5.8  \cr   
76221  & 1338 & X Cnc  & C5,4 & 15&2117 & $-6.0$& 1.13 & 0.35 & 4.8 & binary?\cr
108105 & 1999 & SS Vir & C6,3 & 16&1976 &  2.4  & 3.30 & 0.62 & 6.6 & pseudo-orbit \cr
\hline\cr                                                             
\end{tabular}
}
\end{table*}

\clearpage
\newpage

\begin{table}
\caption[]{References to Tables 1-4}
\begin{tabular}{rl}
\hline
0 & This paper\cr
1 & Udry et al. (1997a)\cr
2 & Udry et al. (1997b)\cr
3 & Carquillat et al. (1997)\cr
4 & Jorissen et al. (1996)\cr
5 & Griffin (1996)\cr
6 & Griffin et al. (1996)\cr
7 & Jorissen et al. (1995)\cr
8 & Jorissen et al. (1993)\cr
9 & Lambert et al. (1993)\cr
10& Jorissen \& Mayor (1992)\cr
11& Griffin \& Keenan (1992)\cr
12& Griffin (1991)\cr
13& McClure \& Woodsworth (1990)\cr
14& Ake \& Johnson (1988)\cr
15& Smith \& Lambert (1988)\cr
16& Smith \& Lambert (1987)\cr
17& Little et al. (1987)\cr
18& Griffin (1984)\cr
19& Schindler et al. (1982)\cr
20& Griffin \& Griffin (1980)\cr
21& McClure et al. (1974)\cr
22& Hackos \& Peery (1968)\cr
23& Fekel et al. (1993)\cr
\hline
\end{tabular}
\end{table}
\end{document}